\documentclass[pre,twocolumn,showpacs]{revtex4}

\usepackage{graphicx}
\usepackage{hyperref}

\usepackage{amsmath,amssymb}
\usepackage{bm,color}
\usepackage{cancel}
\usepackage[normalem]{ulem}
\usepackage{upgreek}

\newcommand{\PR}{Phys.~Rev.~}
\newcommand{\PRE}{Phys.~Rev.~E }
\newcommand{\PRL}{Phys.~Rev.~Lett.~}

\newcommand{\RMP}{Rev.~Mod.~Phys.~}
\newcommand{\EPJE}{Eur.~Phys.~J.~E }
\newcommand{\EPL}{Europhys.~Lett.~}

\newcommand{\st}{\text{s}}
\newcommand{\sig}{\widehat{\bm{\sigma}}}

\newcommand{\vv}{\bm{v}}
\newcommand{\hcs}{\text{HCS}}

\newcommand{\ini}{\text{ini}}

\begin{document}
\title{Memory effect in uniformly heated granular gases}
\author{E. Trizac$^1$ and A.\ Prados$^1,^2$}
\affiliation{$^1$ Universit\'e Paris-Sud, Laboratoire de Physique Th\'eorique et Mod\`eles Statistiques, UMR CNRS 8626,
91405 Orsay, France, EU}
\affiliation{$^2$ F\'{\i}sica Te\'{o}rica, Universidad de Sevilla,
Apartado de Correos 1065, E-41080 Sevilla, Spain, EU}

\date{\today}
\begin{abstract}
  We evidence a Kovacs-like memory effect in a uniformly driven
  granular gas. A system of inelastic hard particles, in the low
  density limit, can reach a non-equilibrium steady state when
  properly forced.  By following a certain protocol for the drive time
  dependence, we prepare the gas in a state where the granular
  temperature coincides with its long time value. The temperature
  subsequently does not remain constant, but exhibits a non-monotonic
  evolution with either a maximum or a minimum, depending on the
  dissipation, and on the protocol.  We present a theoretical analysis
  of this memory effect, at Boltzmann-Fokker-Planck equation level,
  and show that when dissipation exceeds a threshold, the response can
  be coined anomalous.  We find an excellent agreement between the
  analytical predictions and direct Monte Carlo simulations.
\end{abstract}
\maketitle

\section{Introduction}\label{sec1}

A granular material is a system comprising a large number of particles
of macroscopic size, so that the collisions between them are inelastic
and mechanical energy is not conserved.  As consequence, the usual
thermodynamical framework cannot be directly applied to these
systems. Typically, the energy needed to move a grain by one diameter
is many orders of magnitude larger than the thermal energy of the
grain at room temperature, which can be considered irrelevant for all
practical purposes. On the other hand, the concept of \textit{granular
  temperature} is often used in the literature; it is nothing but a
measure of the velocity fluctuations in the system, without being
connected to any notion of thermal equilibrium \cite{JNyB96,bte05}.

We focus here on a low density granular system, which is usually
called a granular gas \cite{PyB03,ByP04}. If no energy is input into
the system, it freely cools (in the sense that its granular
temperature monotonically decreases) and may end up in the homogeneous
cooling state \cite{GyS95,BRyC96,NyE98}, provided instabilities
are circumvented by the choice of a small enough system.  The time
dependence of the system can then solely be encoded in the granular
temperature, which in turn verifies Haff's law \cite{Ha83}. On the
other hand, if there is some mechanism that feeds energy into the
system, it eventually reaches a non-equilibrium steady state in which
energy input by the \textit{thermostat} balances in average the energy
loss due to collisions. To the best of our knowledge, and although
this kind of thermostatted or heated granular fluids have been
extensively investigated
\cite{WM96,NyE98,ENTyP99,MyS00,SyM09,MGyT09,ETB06,Z09,GMyT12,GMyT13,P98a},
no attention has been paid to the possible existence of memory
effects. On the other hand, in other experiments with granular matter
like compaction processes, memory effects have been analyzed both
experimentally and theoretically
\cite{JTMyJ00,ByP01,ByL01,ByP02,RDRNyB05,RRPByD07}.  They have shown
that, in general, the evolution of a compacting granular system
depends not only on the instantaneous value of its packing fraction
but also on its previous history.

A classic experiment in this context is the one performed by Kovacs
fifty years ago \cite{Ko63,Ko79}. A sample of polyvinyl acetate was
equilibrated by putting it in a thermal bath at a high temperature
$T_0$, and then was rapidly quenched to a low temperature $T_1$. At
this low temperature, it relaxed for a given \textit{waiting time}
$t_w$. At time $t=t_w$, the bath temperature was suddenly raised to an
intermediate temperature $T$, $T_0>T>T_1$, such that the instantaneous
value of the polymer volume at $t=t_w$ was equal to its equilibrium
value at $T$.  The behavior of the system for $t > t_w$ was quite
complex: The volume did not remain constant, but increased at first,
passing through a maximum, and relaxed to equilibrium only for longer
times.  As the pressure $P$ was kept fixed along all the process, the
observed behavior means that the knowledge of the state variables $(P,
V ,T)$ does not suffice to completely characterize the state of the
system. The system evolution from an initial state with given values
of $(P,V,T)$ depends on the previous thermal history.  This behavior
is sometimes referred to in the literature as the Kovacs hump, and it
has been extensively studied in glassy and other complex systems
\cite{Br78,ByL01,ByB02,MyS04,AAyN08,PyB10,ByL10,DyH11,RyP14}. {In
  many of these works, the physical quantity displaying the Kovacs
  hump is the energy instead of the volume. In connection with the
  work presented here, it should be emphasized that the granular
  temperature is essentially the internal energy of the granular gas.}
We refer to the driving program in which $T_{1}<T<T_{0}$ as the
``cooling'' protocol. Conversely, a ``heating'' protocol in which the
temperature jumps are reversed and $T_{1}>T>T_{0}$ has been recently
considered \cite{DyH11}. Within this scheme, the relevant physical
quantity, typically the volume or the energy, displays a minimum
instead of a maximum.

{First, it is important to stress that a relevant
  question is the number and type of variables characterizing the
  macroscopic state of granular gases. In the homogeneous cooling
  state \cite{GyS95,BRyC96,NyE98}, and also in the Gaussian
  thermostated case \cite{MyS00,Lu01,BRyM04}, the granular temperature
suffices. For other energy injection mechanisms, like the stochastic
thermostat, there is some evidence that additional variables must be
taken into account: This
uniformly driven granular gas evolves to a hydrodynamic
solution ($\beta$-state) of the kinetic equation
\cite{GMyT12,GMyT13}, over which the granular
temperature is a monotonic function of time. In addition, the granular
temperature and the
driving intensity characterize the $\beta$-state completely, a
behavior that may lead to the conclusion that no  Kovacs
hump should be expected. We show here that this speculative conclusion
is flawed: the Kovacs effect is indeed present in driven granular
gasses and, moreover, it changes sign with inelasticity. }

{In light of the discussion above, it seems worthwhile to
  investigate the possible existence of memory effects in driven
  granular gases.} The steady value of the granular temperature is a
certain function of the driving intensity, which is the externally
controlled parameter in this case. Thus, the granular temperature
plays the role of the volume in the Kovacs experiment, while the
intensity of the driving is the analogue of the bath temperature: we
may start from the stationary state corresponding to a high value of
the driving, and let the system relax to a new steady state by rapidly
quenching the driving to a low value. This relaxation is subsequently
interrupted after a waiting time $t_w$, and the driving is readjusted
to an intermediate value, whose corresponding steady granular
temperature equals its instantaneous value at the waiting time.  The
existence or non-existence of a Kovacs hump in this program
undoubtedly answers whether the granular temperature, together with
the driving intensity, thoroughly characterizes or not the state of
the heated granular system.

In this paper, we investigate the existence of such a hump in the
granular temperature when the above sketched stepwise driving program,
\`a la Kovacs, is implemented in an homogeneously driven granular
gas. We do this analysis both in the usual ``cooling'' protocol (by
decreasing the driving from its initial value) and for the ``heating''
protocol (by increasing the driving from its initial value). In both
cases, we show that the granular temperature indeed displays this
Kovacs hump, thus proving that the granular temperature does not
uniquely characterize the state of the granular system. This is in
agreement with recent investigations in the so-called universal
reference state \cite{GMyT12}, which plays the main role in the
derivation of linear hydrodynamics for driven granular gases
\cite{GMyT13}. However, it will appear that an additional quantity
should be kept in the dynamical description, measuring
non-Gaussianities. Interestingly, there is a value of the restitution
coefficient for which the sign of the hump reverses.  For the cooling
(resp.~heating) protocol, while the granular temperature has a maximum
(resp.~minimum) for high enough restitution coefficient (small
inelasticities), it shows a minimum (resp.~maximum) when the
restitution coefficient is smaller than a critical one (high
inelasticities).  The theoretical results, obtained from the
Boltzmann-Fokker-Planck equation, by (i) considering the first Sonine
approximation and (ii) neglecting nonlinear terms in the excess
kurtosis, are compared to direct Monte Carlo simulations thereof, and
an excellent agreement is found. It is also shown that the expression
of the Kovacs hump so obtained tends to the universal reference state
\cite{GMyT12} for very long times.

The plan of the paper is as follows. In Sec.~\ref{sec2}, we introduce
our model and summarize some of the previous results that are relevant
for the work presented here. In particular, we write the evolution
equations for both the granular temperature and the excess kurtosis of
the velocity distribution function. We put forward a Kovacs-like
program for the driving in Sec.~\ref{sec3}, and obtain approximate
analytical expressions for the time evolution of both the granular
temperature and the excess kurtosis. These analytical expressions are
compared to direct Monte Carlo simulation results.  We
present a physical discussion of the sign and magnitude of
the memory effect in Sec.~\ref{sec:sign_and_mgnitude}. We also discuss
the long time limit and the tendency to the universal reference state
in Sec.~\ref{sec3b}. Some final remarks, relevant to put our work in a
proper context, are presented in Section \ref{sec4}. Preliminary
accounts on parts of this work were published in \cite{PyT14}.


\section{Uniformly heated granular gas}\label{sec2}

We consider a system of $N$ inelastic smooth hard particles of mass
$m$ and diameter $\sigma$. The collisions between them are inelastic
and characterized by the coefficient of normal restitution $\alpha$,
which we assume does not depend on the relative velocity.  In a binary
collision of particles $i$ and $j$, the relation between the
pre-collisional velocities $(\vv_i,\vv_j)$ and the
post-collisional velocities $(\vv'_i,\vv'_j)$ is
\begin{equation}\label{1.1}
\vv'_i=\vv_i-\frac{1+\alpha}{2}\left(\hat{\bm{\sigma}}\cdot\vv_{ij}\right)\bm{\sigma}, \quad \vv'_j=\vv_j+\frac{1+\alpha}{2}\left(\hat{\bm{\sigma}}\cdot\vv_{ij}\right)\bm{\sigma},
\end{equation}
where $\vv_{ij}\equiv\vv_i-\vv_j$ is the relative
velocity and $\sig$ is the unit vector pointing from the
center of particle $j$ to the center of particle $i$ at the collision. Moreover, independent white noise forces act over each grain, so that the following Boltzmann-Fokker-Planck
equation holds for a homogeneous system \cite{NyE98,ENTyP99},
\begin{eqnarray}
  \frac{\partial}{\partial t}f(\vv_1,t)&=&
\sigma^{d-1}\int d\vv_2 \, \bar{T}_0(\vv_1,\vv_2)
f(\vv_1,t)f(\vv_2,t) \nonumber \\
&& +\frac{\xi^2}{2}\frac{\partial^2}{\partial\vv_1^2}f(\vv_1,t), \label{1.2}
\end{eqnarray}
where $d$ is the dimension of space, $\xi$ is a measure of the noise intensity, and $\bar{T}_0$ is the binary collision operator defined by
\begin{equation}\label{1.3}
\bar{T}_0(\vv_1,\vv_2)=\int d\sig \,
\Theta(\vv_{12}\cdot\sig)
(\vv_{12}\cdot\sig)(\alpha^{-2}b_{\sigma}^{-1}-1).
\end{equation}
In the equation above, the operator $b_{\sigma}^{-1}$ replaces the
velocities $\vv_1$ and $\vv_2$ by the precollisional ones, which would be obtained by inverting \eqref{1.1}.
We assume here that the system remains spatially homogeneous, which is backed up by
molecular dynamics simulations \cite{ENTyP99}: the velocity probability
distribution $f$ is thus a sole function of velocity and time.

The granular temperature $T(t)$ is defined as usual,
\begin{equation}\label{1.4}
  n \left\langle\frac{1}{2}m v^2(t)\right\rangle \equiv \int d\vv \frac{1}{2}mv^2 f(\vv,t)=\frac{d}{2} n T(t),
\end{equation}
where $n$ is the density of the system. Moreover, we also introduce the \textit{excess kurtosis} or second Sonine coefficient $a_2$ of the velocity distribution,
\begin{equation}\label{1.5}
  a_2=\frac{d}{d+2} \frac{\langle v^4\rangle}{\langle v^2\rangle^2}-1.
\end{equation}
The excess kurtosis measures the departure from a Gaussian distribution, for which $a_2$ vanishes. It is worth remembering that $\int d\vv f(\vv,t)=n$, so that
\begin{equation}\label{1.6}
  \langle v^n\rangle \equiv \frac{1}{n}\int d\vv \,  v^n f(\vv,t).
\end{equation}
Starting from the Boltzmann-Fokker-Planck equation \eqref{1.2}, one can derive the equation governing the time evolution of the granular temperature
\begin{equation}\label{1.7}
  \frac{dT}{dt}=m \xi^2 - \zeta_0 T^{3/2} \left( 1+\frac{3}{16}a_2 \right),
\end{equation}
where
\begin{equation}\label{1.8}
  \zeta_0=\frac{2 n \sigma^{d-1} \left(1-\alpha^2\right) \pi^{\frac{d-1}{2}}}{\sqrt{m}\, d\,\Gamma(d/2)}.
\end{equation}
Equation \eqref{1.7} is valid in the so-called first Sonine approximation, and terms of $\mathcal{O}(a_2^2)$ are neglected in its derivation \cite{NyE98}
together with higher order contributions, that do not seem to be relevant \cite{BP06}.
In other words, the velocity distribution is expanded in the form,
\begin{eqnarray}
f(\vv,t) \,=\, \frac{ e^{-v^2/v_0^2}}{v_0^d \, \pi^{d/2}} \,\left[
1+ a_2 \, S_{2}(v/v_{0})\right] , \\
S_{2}(x)=\frac{1}{2}\,x^4 -\,\frac{d+2}{2} x^{2} +\frac{d(d+2)}{8}
\end{eqnarray}
where $v_0$ is the time dependent typical velocity defined by
$T=mv_0^2/2$, and $S_{2}(x)$ is the second Sonine polynomial.
{Sonine-related techniques are often useful in kinetic theory \cite{Landau},
to study the non equilibrium behaviour of dissipative gases \cite{ByP02} or in the context
of ballistically controlled irreversible dynamics
\cite{T02,PTD02}}.

In the
long time limit, the system approaches a steady state in which the
energy input due to the white noise force balances on average the
energy loss due to the collisions. Therefore, the granular temperature
$T$ and the excess kurtosis $a_2$ approach their steady values $T_\st$
and $a_2^\st$, respectively, which verify
\begin{equation}\label{1.9}
  m\xi^2=\zeta_0 T_\st^{3/2} \left( 1+\frac{3}{16}a_2^\st \right).
\end{equation}
The evolution equation \eqref{1.7} or its particularization to the steady state \eqref{1.9} are not closed for the granular temperature, because of the terms proportional to the
excess kurtosis in them.  The steady value of the excess kurtosis can be calculated in the first Sonine approximation \cite{NyE98,SyM09}
\begin{equation}\label{1.10}
a_2^\st=\frac{16(1-\alpha)(1-2\alpha^2)}
{73+56d-24d\alpha-105\alpha+30(1-\alpha)\alpha^2}.
\end{equation}
Then, the steady value of the temperature is
\begin{equation}\label{1.11}
T_\st=m\left[\frac{d\Gamma(d/2)\xi^2}
{2\pi^{\frac{d-1}{2}}n\sigma^{d-1}(1-\alpha^2)(1+\frac{3}{16}a_2^\st)}\right]^{2/3}.
\end{equation}
Let us turn Eq. \eqref{1.7} into an evolution equation for the dimensionless variable
 \begin{equation}\label{1.12}
   \beta=\sqrt{\frac{T_\st}{T}}
 \end{equation}
that measures the separation of the temperature from its steady value. A simple calculation yields
\begin{equation}\label{1.13}
  \frac{d\beta}{dt}=\frac{\zeta_0}{2} \sqrt{T_\st} \left[ 1+\frac{3}{16}a_2-\left(1+\frac{3}{16}a_2^\st \right) \beta^3 \right].
\end{equation}

The evolution equation for the excess kurtosis can also be derived from the Boltzmann-Fokker-Planck equation \cite{GMyT12}. We again consider the first Sonine approximation and
neglect nonlinear terms in the excess kurtosis, to obtain that
\begin{equation}\label{1.14}
  \beta \frac{da_2}{dt}= 2\zeta_0 \sqrt{T_\st} \left[ \left(1-\beta^3\right) a_2+B \left( a_2^\st -a_2 \right) \right].
\end{equation}
The parameter $B$ has been computed in \cite{GMyT12,typo}, with the result
\begin{widetext}
\begin{equation}\label{1.15}
  B=\frac{73+8d(7-3\alpha)+15\alpha[2\alpha(1-\alpha)-7]}
{16(1-\alpha)(3+2d+2\alpha^2)
+a_2^\st[85+d(62-30\alpha)+3\alpha(10\alpha(1-\alpha)-39)]},
\end{equation}
\end{widetext}
which is then a given function of the restitution coefficient and of the dimension of space.
It turns out, however, that it can be obtained from a self-consistent argument \cite{PyT14}.
In the limit where the forcing $\xi$ is so small that $\beta\to 0$, the excess kurtosis
should evolve to its homogeneous cooling state value, given by \cite{SyM09}
\begin{equation}
a_2^\hcs=\frac{16(1-\alpha)(1-2\alpha^2)}{25 +2\alpha(\alpha-1) + 24 d + \alpha ( 8 d-57 )}.
\label{eq:a2hcs}
\end{equation}
This yields a strong constraint on $B$, which has to be compatible with this
requirement. In other words, the right hand side of Eq. (\ref{1.14}),
when $\beta$ can be neglected, should admit $a_2^\hcs$ as a root.
Thus,
\begin{equation}
 a_2^\hcs+B \left( a_2^\st -a_2^\hcs \right)= 0
\end{equation}
from which we obtain that
\begin{eqnarray}
B &=& \frac{a_2^\hcs}{a_2^\hcs-a_2^\st} \\
&=&\frac{73+8d(7-3\alpha)+15\alpha[2\alpha(1-\alpha)-7]}
{16(1-\alpha)(3+2d+2\alpha^2)} .
\label{eq:B}
\end{eqnarray}
This expression, interestingly, is derived in a more straightforward
way than in Ref. \cite{GMyT12}. {They differ by the the term
  proportional to $a_{2}^{\st}$ in the denominator of
  Eq.~\eqref{1.15}, which reduces to Eq.~\eqref{eq:B} if this term is
  omitted.} In the following analysis, we will make use of
Eq.~\eqref{eq:B} instead of Eq.~\eqref{1.15}, since it turns out to be
more accurate as compared to simulation results. In addition, {this} is
consistent with the linearization in $a_{2}$ in Eq.~\eqref{1.14}:
Therein, $B$ multiplies $a_{2}-a_{2}^{\st}$, so that any terms
proportional to the excess kurtosis in $B$ should be neglected.

Equation \eqref{1.14}, together with \eqref{1.13}, constitute a closed
set of two differential equations for the time evolution of the
rescaled temperature $\beta$ and the excess kurtosis $a_2$. We can
also introduce a rescaled excess kurtosis
\begin{equation}\label{1.16}
  A_2=\frac{a_2}{a_2^\st}, \quad A_2^\st=1,
\end{equation}
and rewrite  Eqs. \eqref{1.13} and \eqref{1.14} in the following way,
\begin{subequations}\label{1.17}
\begin{equation}\label{1.17a}
  \frac{d\beta}{d\uptau} \,=\, 1-\beta^3 +\frac{3}{16}a_2^\st \left( A_2-\beta^3 \right),
\end{equation}
\begin{equation}\label{1.17b}
  \beta \frac{dA_2}{d\uptau} \,=\, 4 \left[ \left(1-\beta^3\right) A_2+ B \left( 1-A_2 \right) \right],
\end{equation}
\end{subequations}
where we have introduced a rescaled time
\begin{equation}\label{1.18}
  \uptau \, = \, \frac{\zeta_0 \sqrt{T_\st}}{2} \, t.
\end{equation}
Equations \eqref{1.17} are nonlinear in $\beta$ but linear in the excess kurtosis, consistently with our approach. Obviously, $\beta=1$ and $A_2=1$ is a stationary solution.

\section{Memory effect}\label{sec3}

\begin{figure}
  \centering
  \includegraphics[width=3.25in]{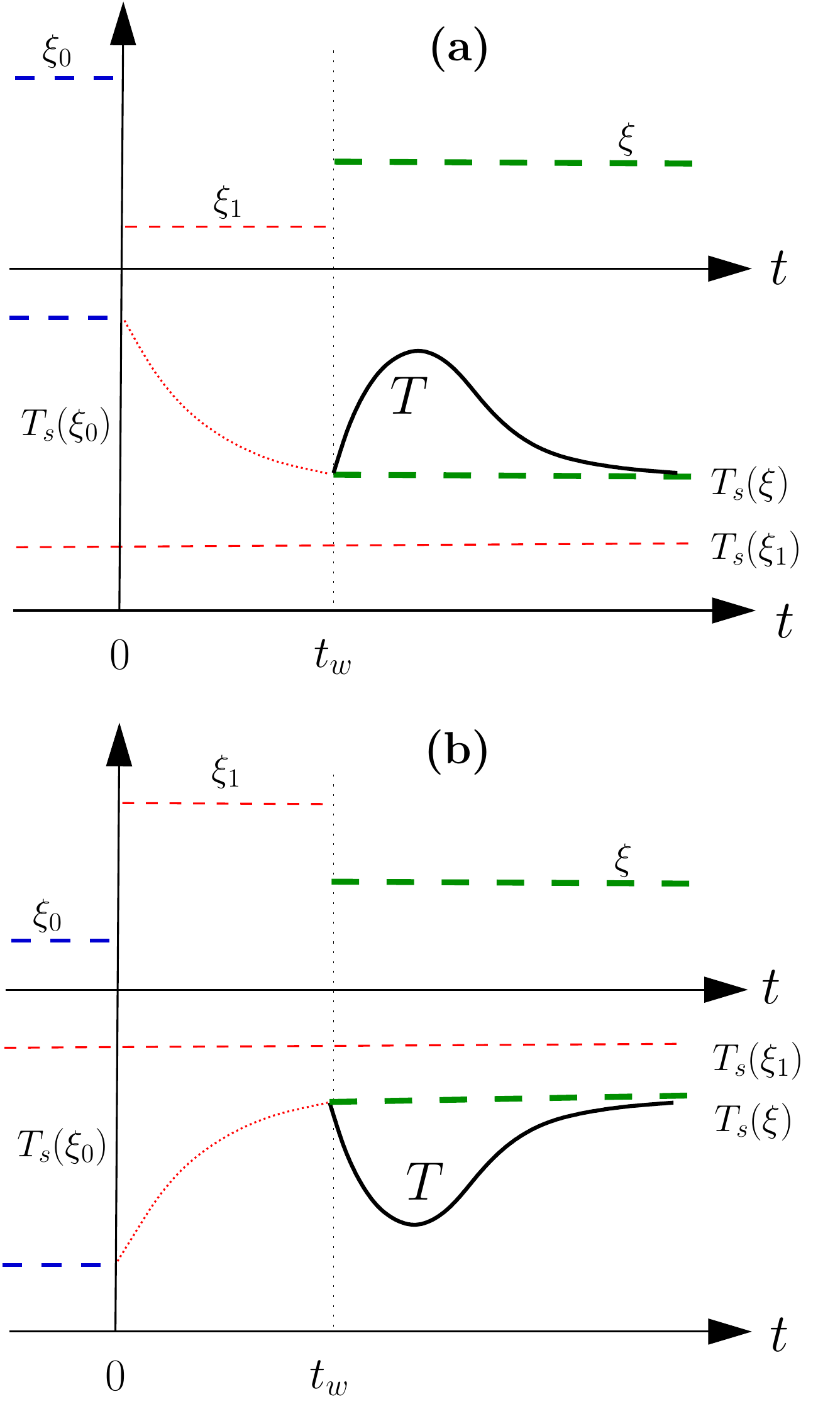}\\
  \caption{Sketch of the drive time dependence for the cooling and
    heated protocols.  The resulting {\em normal} temperature
    evolution is depicted. The system is first in a non-equilibrium
    steady state at temperature $T_s(\xi_0$) under a drive $\xi_0$.
    $T(t_w)$ coincides with $T_s(\xi)$.  (a) Cooling protocol:
      The driving $\xi_{1}$ in the waiting time window $0<t<t_{w}$ is
      smaller than its initial value $\xi_{0}$, and the granular
      temperature would display a maximum before returning to its
      steady value for $t>t_{w}$. (b) Heating protocol: We have that
      $\xi_{1}>\xi_{0}$ and the granular temperature would display a
      minimum for $t>t_{w}$. }
  \label{fig:protocol}
\end{figure}

We are interested in analyzing the following experiment. First, we let
a system of inelastic hard particles reach the steady state
corresponding to some value of the driving, say $\xi_0$. Then, at
$t=0$ we quench the driving to either $\xi_1 <\xi_0$ (cooling
protocol), or to $\xi_1 >\xi_0$ (heating protocol), and the system
subsequently evolves for a time $t_w$, the \textit{waiting time}. At
$t=t_w$, we measure the granular temperature and suddenly change the
driving to the value $\xi$ such that the stationary granular
temperature $T_\st(\xi)$ equals the measured value at $t_w$,
$T(t=t_w)$. This amounts to $\xi_1<\xi<\xi_0$ in the cooling case, and
$\xi_1>\xi>\xi_0$ in the heated one, see Fig. \ref{fig:protocol}.  If
the state of the system were completely determined by the granular
temperature, as is the case in the homogeneous cooling state, the temperature would remain
constant for $t>t_w$. But, since the values of the excess kurtosis for
$t=t_w$ and for the steady state corresponding to the final driving
$\xi$ are different, the granular temperature will separate from its
steady value at first, pass through an extremum, and only return to
its steady (initial) value for longer times. We may refer to this
behavior as the Kovacs hump, because it is similar to the so-called
behavior in polymers, structural glasses and other complex systems
\cite{Ko63,Ko79,Br78,ByB02,MyS04,AAyN08,PyB10,ByL10,DyH11,RyP14}.

In the analogous experimental situation for molecular systems, when
the ``driving'' is first lowered ($\xi_0\to\xi_1$) and afterwards
increased to an intermediate value ($\xi_1\to\xi<\xi_0$), the measured
quantity, typically the volume \cite{Ko63,Ko79,MyS04,ByL10} or the
energy \cite{Br78,ByB02,AAyN08,PyB10,DyH11,RyP14}, always passes
through a maximum. An analogous behavior is expected for any
  physical quantity that increases with increasing temperature. On
the other hand, within the heated protocol, a minimum is expected,
as theoretically predicted by linear response theory
  \cite{PyB10}. Moreover, in the nonlinear regime, the existence of
  this minimum for the heated protocol has been recently checked for a
  simple model \cite{DyH11}.  We will refer to this behavior,
in which the time derivative of the energy changes sign at
  $t_{w}$, {that is, the energy displays a rebound}, as `normal'.  {It} must be stressed here that the final
state of the granular gas is not an equilibrium one, but an
out-of-equilibrium stationary state, and thus the behavior of the
granular temperature may be different.

\subsection{Analytical results}
\label{ssec:analytical}

The evolutions in the waiting window ($0\leq t\leq t_w$), and for
$t\geq t_w$ both obey the differential equations \eqref{1.17}, but
with different initial conditions.  At $t=0$, we have $A_2=1$ with
either $\beta<1$ (cooling protocol) or $\beta>1$ (heating protocol).
At $t=t_w$, a 'reversed' condition should be enforced, with $\beta=1$
while $A_2$ results from the dynamics in the waiting
window. $A_2(t_w)$ turns out to be larger than 1 for the cooling
protocol, and smaller than 1 in the heated case (see Sec.~\ref{ssec:optimaltw}). Since the waiting
time dynamics only enters through the value of $A_2(t_w)$, we assume
the latter given, and concentrate on the evolution at $t>t_w$.  We
shall use the rescaled time $\uptau$ introduced in \eqref{1.18}, with
$\uptau_w = \zeta_0 \sqrt{T_\st} \, t_w/2$.

 \begin{figure}
  \centering
  \includegraphics[width=3.25in]{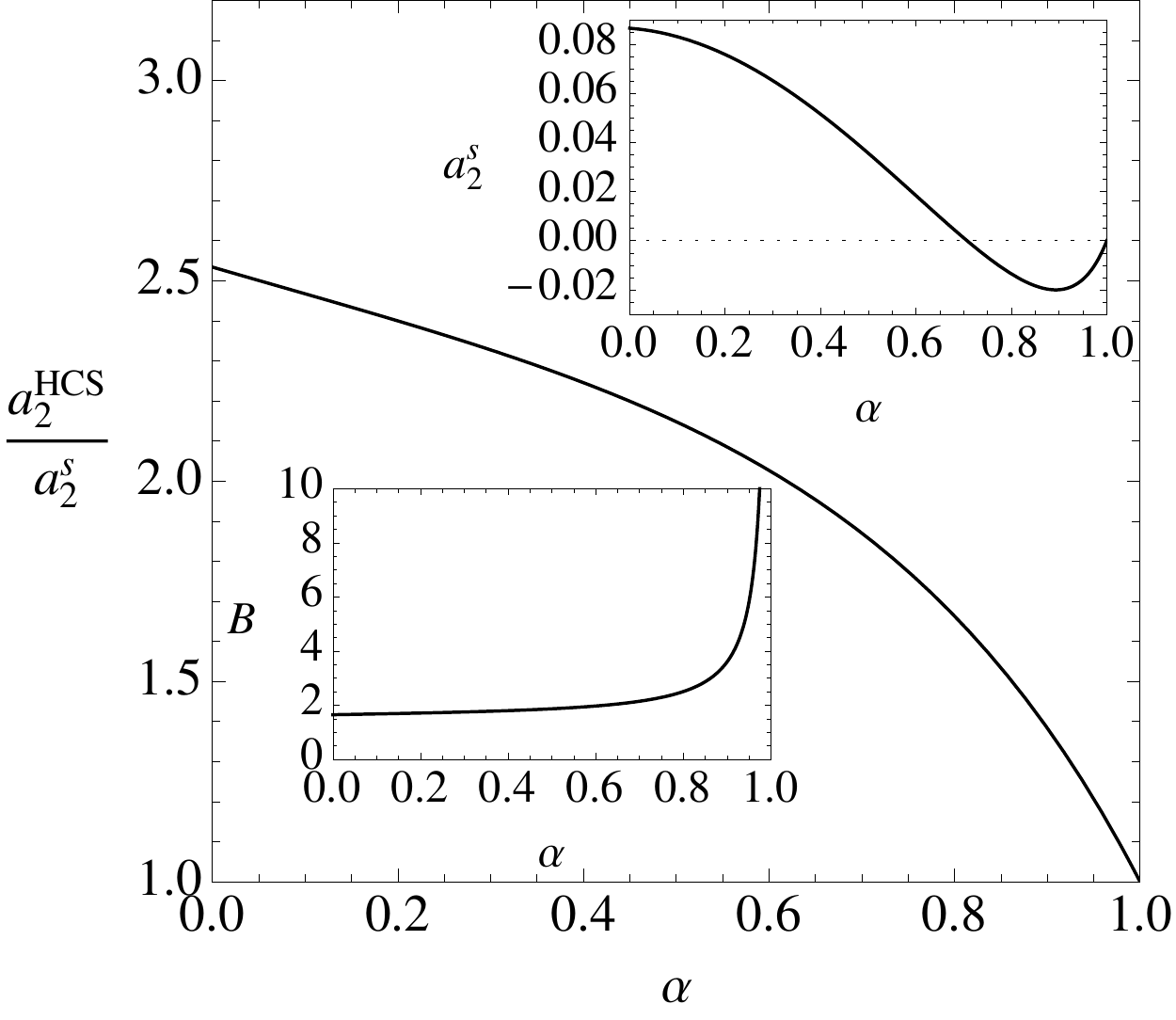}\\
  \caption{Plot of $ a_2^\hcs/a_2^\st$ as a function of the
    restitution coefficient $\alpha$, for a system of inelastic hard
    disks ($d=2$), following from the accurate expressions obtained in
    \cite{SyM09}. The top and bottom insets show the excess kurtosis
    for the steady state $a_{2}^{\st}$ and the parameter $B$ as
    functions of $\alpha$, as given by Eq.~\eqref{1.10} and
    \eqref{1.15}, respectively.  }
  \label{fig:a2ratio}
\end{figure}

Equations \eqref{1.17} with the initial conditions
\begin{equation}\label{2.2}
  \beta(\uptau=\uptau_w)=1, \quad A_2(\uptau=\uptau_w)
  \equiv A_2^{\ini},
\end{equation}
do not seem to admit an analytical solution, but an approximate and
accurate method can be found in the following way.  The initial value
of $A_2$ is of the order of unity: In the cooling case, $A_{2}$
  is bounded from above by $a_2^\hcs/a_2^\st$, shown in Fig. \ref{fig:a2ratio}
  and, in the heated case, we have that $0<A_2^\ini<1$, as shown in
  Sec.~\ref{ssec:optimaltw} below.  The idea is next to
expand both $\beta$ and $A_2$ in powers of $a_2^\st$. The rationale
for this expansion is the smallness of $a_2^\st$ throughout the whole
inelasticity range, namely $|a_2^\st|\leq 0.086$. Thus we introduce the
series expansions
\begin{subequations}\label{2.5}
\begin{equation}\label{2.5a}
  \beta(\uptau)=\beta_0(\uptau)+a_2^\st \beta_1(\uptau)+\ldots,
\end{equation}
\begin{equation}\label{2.5b}
  A_2(\uptau)=A_{20}(\uptau)+a_2^\st A_{21}(\uptau)+\ldots,
\end{equation}
\end{subequations}
into \eqref{1.17}, and write the subsequent equations up to linear
order in $a_2^\st$. To the zero-th order we have
\begin{equation}\label{2.6}
  \frac{d\beta_0}{d\uptau}=1-\beta_0^3, \quad \beta_0 \frac{dA_{20}}{d\uptau}=4 \left[(1-\beta_0^3)A_{20}+B(1-A_{20})\right],
\end{equation}
submitted to the initial conditions $\beta_0(\uptau=\uptau_w)=1$ and $A_{20}(\uptau=\uptau_w) = A_2^\ini$. Therefore, $\beta_0(\uptau)=1$, $\forall \uptau$,
\begin{equation}\label{2.7}
   \quad \frac{dA_{20}}{d\uptau}=-4B\left(A_{20}-1\right).
\end{equation}
The zero-th order solution is then
\begin{subequations}\label{2.8}
\begin{equation}\label{2.8a}
  \beta_0(\uptau)=1,
\end{equation}
\begin{equation}\label{2.8b}
  \quad A_{20}(\uptau)=1+\Delta A_2^\ini e^{-4B(\uptau-\uptau_w)}, \quad \Delta A_2^\ini\equiv A_2^\ini-1.
\end{equation}
\end{subequations}
To this order, the granular temperature $\beta_0$ remains constant while $A_{20}$ relaxes exponentially from its initial to its steady state value with a characteristic time (in the $\uptau$ scale)
\begin{equation}\label{2.9}
  \uptau_c=(4B)^{-1}.
\end{equation}
There is consequently no memory effect to zeroth order.

The equation for the first order contribution to the scaled temperature is
\begin{equation}\label{2.10}
  \frac{d\beta_1}{d\uptau}=-3\beta_1 +\frac{3}{16} \Delta A_2^\ini e^{-4B(\uptau-\uptau_w)}, \quad \beta_1(\uptau=\uptau_w)=0,
\end{equation}
whose solution is readily obtained as
\begin{equation}\label{2.11}
  \beta_1(\uptau)=\gamma \Delta A_2^\ini \left( e^{-3(\uptau-\uptau_w)}-e^{-4B(\uptau-\uptau_w)} \right).
\end{equation}
We have introduced the definition
\begin{equation}\label{2.12}
  \gamma=\frac{3}{16(4B-3)}>0,
\end{equation}
which is positive definite because $B>3/4$, see
Fig. \ref{fig:a2ratio}. The parameter $\gamma$ depends on the
restitution coefficient $\alpha$ and the dimension of space $d$, as
does $B$. {Note that we have only needed the zero-th order
  approximation $A_{20}$ for calculating the evolution of the
  temperature up to first-order in the perturbation parameter
  $a_{2}^{\st}$, that is, $\beta_{1}$. This stems from the
  mathematical structure of the equation for $\beta$ in \eqref{1.17},
  in which $A_{2}$ only appears in the term proportional to
  $a_{2}^{\st}$. We will consider the first-order correction $A_{21}$
  to the excess kurtosis in Sec.~\ref{sec3b}, in connection with the
  long time behavior of the solution.}

{Equation \eqref{2.11} implies} that
the sign of $\beta_1(\uptau)$ is the same as the sign of $A_2^\ini
-1$, which can be shown to be positive for the cooling procedure, and
negative in the heated case. We will come back to this feature in
Sec.~\ref{ssec:optimaltw}.  The time evolution for the temperature,
obtained by substituting \eqref{2.8a} and \eqref{2.11} into
\eqref{2.5a}, is given by
\begin{eqnarray}
  \beta(\uptau)-1 & = & a_2^\st \gamma \Delta A_2^\ini \left( e^{-3(\uptau-\uptau_w)}-e^{-4B(\uptau-\uptau_w)} \right)  \nonumber \\
  &=& \gamma \left(a_2^\ini-a_2^\st \right) \left( e^{-3(\uptau-\uptau_w)}-e^{-4B(\uptau-\uptau_w)} \right) , \nonumber \\
  && \label{2.13}
\end{eqnarray}
up to higher order terms in $\mathcal{O}(a_2^\st)^2$.
Thus, the sign of the ``distance'' $\beta-1$ of the granular temperature to its steady value  is the same as that of
$(a_2^\ini-a_2^\st )$. If $\alpha$ is changed, it affects both $a_2^\st$
and $a_2^\ini$ so that $(a_2^\ini-a_2^\st )$ and $a_2^\st$ share the same sign, which changes at a certain value of the restitution coefficient,
$\alpha_c \simeq 1/\sqrt{2}\simeq 0.707$ \cite{rque88}: as a consequence,
$a_2^\st>0$ for $\alpha<\alpha_c$ while $a_2^\st<0$ for
$\alpha>\alpha_c$, see the top inset in Fig.~\ref{fig:a2ratio}.
We now restrict the discussion to cooling protocols. The above reasoning implies that for high inelasticities,
namely $\alpha<\alpha_c$, $\beta-1>0$ and then $\beta$ has a maximum while the granular temperature has a minimum (remember that $T=T_\st/\beta^2$). The situation reverses
for small inelasticities, $\alpha>\alpha_c$, for which $\beta-1<0$. Then, $\beta$  has a minimum,  which corresponds to a maximum of the granular temperature.
On the other hand, for heating protocols, the phenomenology is reversed, but ruled by very similar
mechanisms. For $\alpha>\alpha_c$, $T$ shows a minimum, whereas for $\alpha<\alpha_c$, it exhibits
a maximum. A more physical explanation will be provided in subsection
\ref{ssec:physical_mechanism}.

It should be noted here that from the structure of Eq. \eqref{2.13}, the shape of the hump (the $\uptau$ dependence) and
its amplitude are factorized. In other words, Eq.~\eqref{2.13}
  can be rewritten as
\begin{subequations}\label{factorized}
\begin{eqnarray}
  \beta(\uptau)-1 &= & g(\uptau_{w})\, h(\uptau-\uptau_{w}),
  \label{eq:factorized1} \\
  g(\uptau_{w})&=&a_{2}^{\st}\Delta A_{2}^{\ini}=a_{2}^{\ini}-a_{2}^{\st},  \label{eq:factorized2}\\
 h(s)&=& \gamma  \left(
   e^{-3s}-e^{-4Bs} \right)>0.
\label{eq:factorized3}
\end{eqnarray}
\end{subequations}
The prefactor $g(\uptau_{w})$ contains all the information about the
details of the protocol in the waiting time window, that is, the
dependence of the hump not only on $t_{w}$ but also on
$\{\xi_{0},\xi_{1}\}$, while $h(\uptau-\uptau_{w})$ determines its
shape. We shall show in Sec.~\ref{ssec:optimaltw} that $\Delta
A_{2}^{\ini}$ has a definite sign for both cooling and heating
protocols, so that $g$ also determines the sign of the hump through
the steady value of the excess kurtosis $a_{2}^{\st}$ or,
equivalently, $a_{2}^{\ini}-a_{2}^{\st}$.

Equation \eqref{2.13} or \eqref{factorized} gives then the lowest
order expression for the Kovacs hump, within the theoretical framework
we have just developed. It clearly shows that the granular temperature
is not enough for describing the state of uniformly heated granular
gases, as has been already claimed by other means
\cite{GMyT12,GMyT13}.  If that were the case, no hump at all would be
present when the system is prepared with the \textit{correct} initial
granular temperature for the subsequent driving, within our \`a la
Kovacs program. On the other hand, the existence of the Kovacs hump
does not directly follow from the non-Maxwellian character of the
velocity distribution. Indeed, although the velocity distribution of a
granular gas is generically non-Gaussian, the granular temperature may
completely specify its state in some situations. This is the case for
the homogeneous cooling state but also for the equivalent system
driven by the so-called Gaussian thermostat. Therein, particles are
accelerated between collisions by a force proportional to their own
velocity \cite{MyS00,Lu01,BRyM04}, and no Kovacs hump would be
observed if an analogous stepwise driving procedure were followed.

\subsection{Numerical results}
\label{sec3a}

We compare here the analytical expression for the Kovacs hump to the
results obtained by direct Monte Carlo simulations \cite{Bi94} of the
Boltzmann-Fokker-Planck equation. We have used a system of $N=10^4$
hard disks ($d=2$) of unit mass, $m=1$, and unit diameter, $\sigma=1$,
with the collision rule \eqref{1.1}. The results have been averaged
over a large number (ranging from $N_T=10^5$ to $1.5\times 10^6$) of
realizations of the stochastic dynamics of the system. The stochastic
thermostat is taken into account by the procedure first introduced in
Ref.~\cite{ENTyP99}. Over each trajectory, the hard disks are
submitted to random kicks every $N_c=N/10^3=10$ collisions. In the
kick, each component of the velocity of every particle is incremented
by a random number extracted from a gaussian distribution of variance
$\xi^2\Delta t$, where $\Delta t$ is the time interval corresponding
to the number of collisions $N_c$. Moreover, every $N/10^2=100$
collisions, a possible non-vanishing center of mass velocity is
eliminated to enforce conservation of momentum and avoid a spurious
drift of the center-of-mass velocity.

\begin{figure}
  \centering
  \includegraphics[width=3.25in]{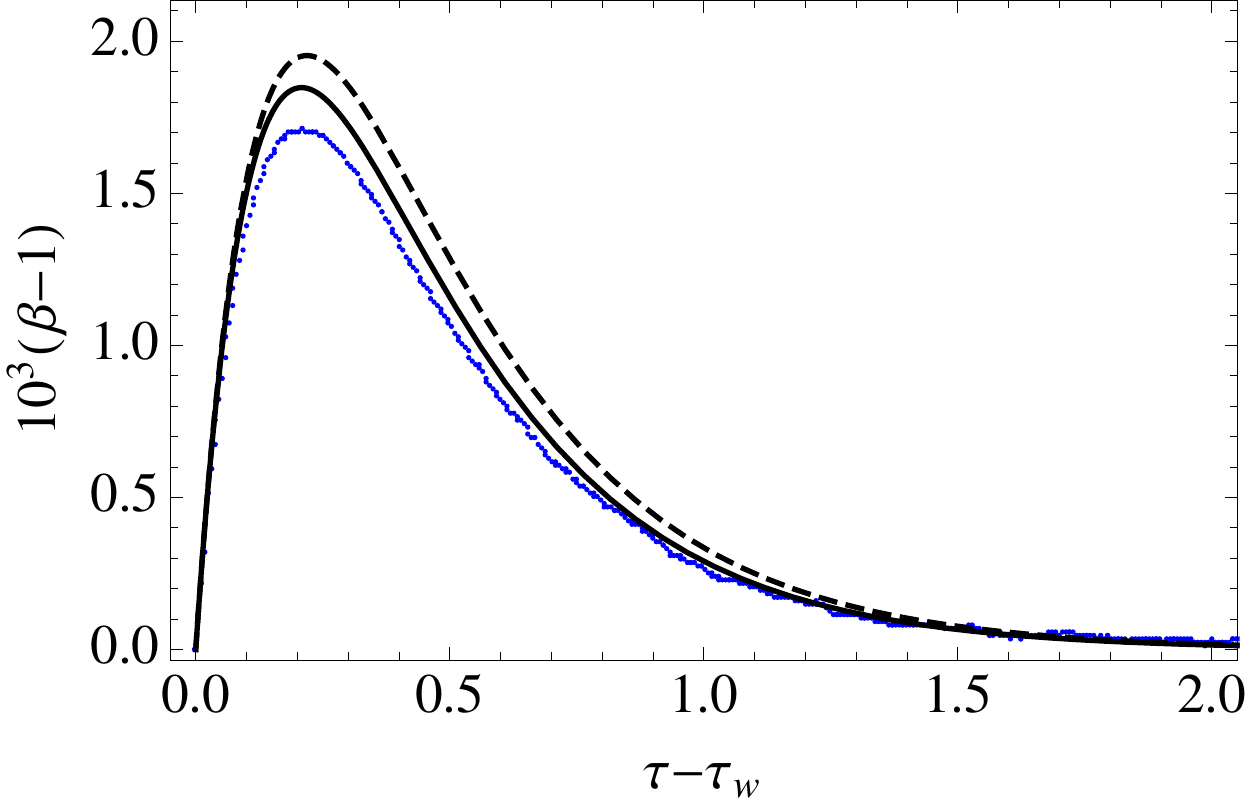}\\
  \includegraphics[width=3.25in]{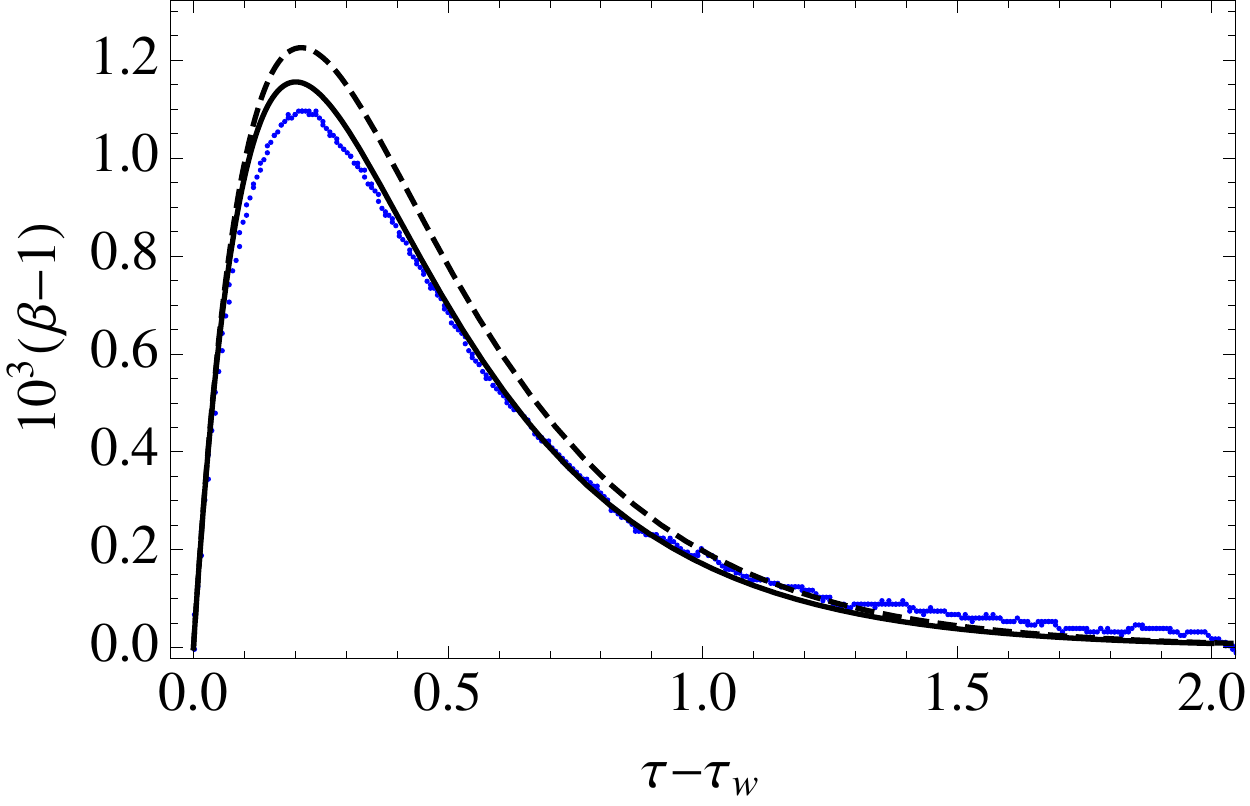}\\
  \caption{Plot of the Kovacs hump for $\alpha=0$ (top) and $\alpha=0.3$ (bottom). The simulation curves (points) have been averaged over $10^5$ trajectories, and they are compared
  to (i) the \textit{raw} theoretical curve \eqref{2.13}, evaluated with the theoretical expressions for the parameters $a_2^\st$, $B$, and $a_2^\hcs$ (dashed line) and (ii) the
  improved theory obtained by inserting into \eqref{2.13} the value of the $B$-parameter given by the Monte Carlo simulation (solid line). The second route improves the agreement
  between theory and simulation. The specific values of the parameters
  for each of the plotted curves are given in Table \ref{table1}. Note
  the smallness of $\beta-1$, which is of the order of $10^{-3}$ in
  both cases.}\label{fig2}
\end{figure}

\begin{table}
  \centering
  \begin{tabular}{|c|c|c|c|}
    \hline
     & $\alpha=0$ & $\alpha=0.3$ & $\alpha=0.8$ \\ \hline
    $B$ from DSMC & 1.802 & 1.920 & 2.440 \\
    $B$ from \eqref{1.15} & 1.422 & 1.555 & 2.602 \\
    $B$ from \eqref{eq:B} & 1.652 & 1.753 & 2.507 \\
    \hline
  \end{tabular}
  \caption{Values of the excess kurtosis decay rate $B$, corresponding
    to the plots in Figs. \ref{fig2} and \ref{fig3}. For comparison
    with  Monte Carlo data, Eq. \eqref{eq:B} has been used.}
  \label{table1}
\end{table}

Our analytical predictions reveal that the Kovacs effect is all the
more pronounced as the difference $|a_2^\ini-a_2^\st|$ is large. Quite
intuitively, there are two ways to maximize $|a_2^\ini-a_2^\st|$:
either taking $\xi_1 \ll \xi_0$ (equivalently $T_s(\xi_1)\ll
T_s(\xi_0)$ in the cooling case, or in the heated situation, reversing
all inequalities. We concentrate here on the cooling protocol, for
which we have performed simulations such that the choice $\xi_1 \ll
\xi_0$ guaranties that the system, in the waiting time window, has an
excess kurtosis that quickly evolves towards its free cooling
counterpart; thus, $A_2(\uptau_w) = a_2^\hcs/a_2^\st$.  We will
discuss in subsection \ref{ssec:optimaltw} the cases of finite $\xi_1
/ \xi_0$.  {For the sake of simplicity, we have always used $\xi_{1}=0$, which allows us to simplify
the simulation procedure, see below.}

{Let us explain how we calculate in the simulations the final
  value of the driving $\xi$ from the value of the granular
  temperature $T(t_{w})$ at the end of the waiting time window. For an
  arbitrary value of the intermediate driving $\xi_{1}$:
  (i) run all the realizations until the waiting
  time, (ii) obtain the granular temperature $T(t_{w})$ averaging over
  all the realizations, (iii) determine the final value of the driving
  $\xi$ therefrom, and (iv) continue running all the
  realizations. This numerical procedure introduces some (in general
  unavoidable) numerical errors, stemming from the fluctuations of the
  granular temperature over the different realizations. Nevertheless,
  we may take advantage of the value of the driving in the waiting
  time window, $\xi_{1}=0$, to eliminate these fluctuations and
  minimize the numerical error. For long enough waiting times
  \cite{long_enough_tw}, the system cools in the homogenous cooling
  state, a regime where all the
  time evolution may be encoded in the granular temperature. Then, we
  proceed in the following way: (i) We choose a value of the final
  driving $\xi$, and calculate the corresponding steady granular
  temperature $T_{\st}(\xi)$, (ii) run each realization until the
  shortest time $t$ such that $T(t)<T_{\st}(\xi)$, (iii) rescale all
  the velocities of the particles with a factor
  $\sqrt{T_{\st}(\xi)/T(t)}$, so that $T(t)=T_{\st}(\xi)$, thus
  effectively eliminating the granular temperature fluctuations at the
  waiting time, and (iv) continue running all the realizations.}

\begin{figure}
  \centering
  \includegraphics[width=3.25in]{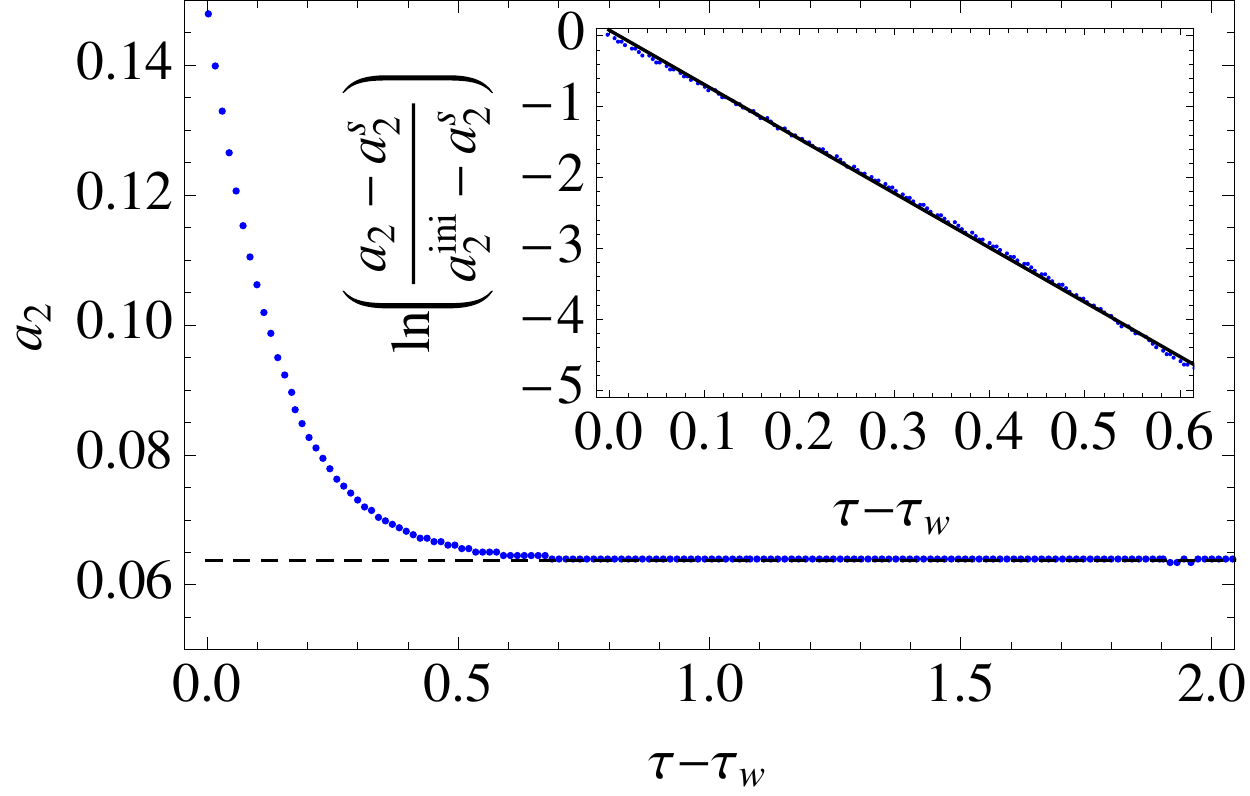}\\
  \caption{Decay of the excess kurtosis from its initial to its steady state value. Plotted is the simulation curve obtained by DSMC (points) for $\alpha=0.3$.
  The long time limit is very close to its predicted value $a_2^\st=0.00638$,
  following from Eq. (\ref{1.10}) and shown by the dashed line.
  In the inset, the same decay but on a logarithmic scale (points). From the linear slope, we directly measure the parameter $B$, to be inserted into the theoretical expression for the
  Kovacs hump, Eq. \eqref{2.13}. The obtained values are given in Table \ref{table1}.}\label{fig:a2decay}
\end{figure}

In Fig. \ref{fig2}, we show the comparison between the numerical
computation of the Kovacs hump and our theoretical prediction, in the
high inelasticity regime $\alpha<\alpha_c\simeq 0.707$.  Namely, we
have considered (a) $\alpha=0$ and (b) $\alpha=0.3$. In both cases,
there are two theoretical curves: the dashed line corresponds to the
raw evaluation of Eq. \eqref{2.13} with the theoretical values of
$a_2^\st$, $a_2^\hcs$ and $B$ given by Eqs. \eqref{1.10},
\eqref{eq:a2hcs} and \eqref{eq:B}, respectively. Although the
qualitative agreement is reasonable, there are quantitative
discrepancies. This is not surprising. While the analytical
predictions for $a_2^\st$ and $a_2^\hcs$ turn out reliable for our
purposes, Eq. \eqref{eq:B} does not fare as well, and may be
  plagued by nonlinear effects, as is the case for Eq. (\ref{1.15})
  \cite{GMyT12}. Therefore, we have followed an alternative route: We
first measure $B$ from the relaxation of the excess kurtosis, as
embodied in relation \eqref{2.8b}, see Fig.~\ref{fig:a2decay}, which
clearly exhibits an exponential behavior.  The corresponding value of
$B$ is then inserted in Eq. \eqref{2.13}, to give the solid line in
Fig. \ref{fig2}. A posteriori, we have also compared the values of $B$
to their analytical counterparts, as seen in Table
\ref{table1}. The inaccuracy of the theoretical estimate is of
  approximately $10\%$ for Eq. \eqref{eq:B}, and 20\% with
  Eq. \eqref{1.15}, consistently with the situation found in previous
  studies \cite{GMyT12}. It appears that once an accurate value of
the relaxation parameter $B$ is known, quantitative predictions can be
made.

\begin{figure}
  \centering
  \includegraphics[width=3.25in]{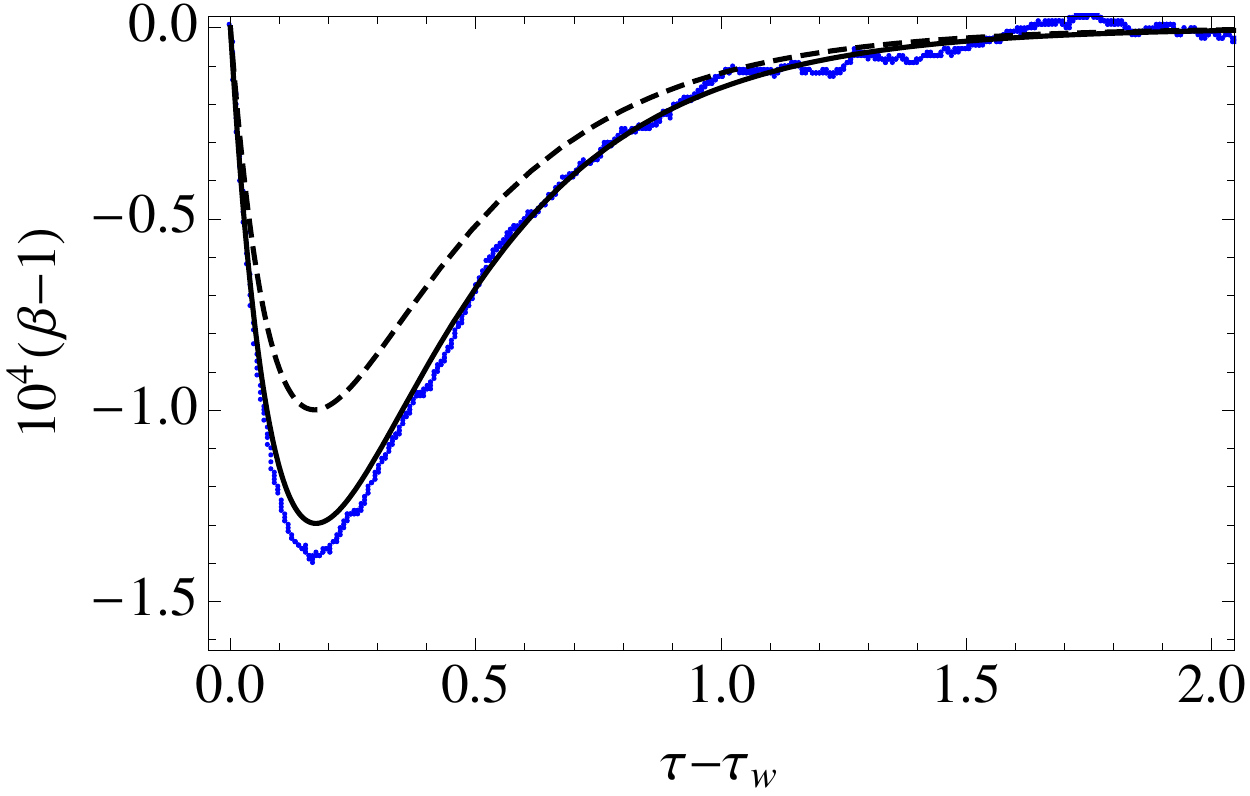}\\
  \caption{Plot of the Kovacs hump for $\alpha=0.8$. The meaning of the different symbols and lines is the same as in Fig.~\ref{fig2}. Note that the sign of $\beta-1$ is reversed, $\beta-1<0$
  as the restitution coefficient $\alpha>\alpha_c\simeq 0.707$.}\label{fig3}
\end{figure}

Figure \ref{fig3} shows the Kovacs hump for a smaller value of the inelasticity, namely $\alpha=0.8>\alpha_c$. As predicted by the theory, the sign of $\beta-1$ is reversed, since $a_2^\st<0$
for $\alpha>\alpha_c$. The simulation curve has been averaged over $1.5\times 10^6$ trajectories, because in this region not only $|a_2^\st|$ but also $\Delta A_2^\ini$ are of  smaller magnitude,
see Fig.~\ref{fig:a2ratio}. Thus, the amplitude of the hump is reduced roughly tenfold as compared to those in Fig.~\ref{fig2}. For $\alpha=0.8$, the error in the theoretical estimate of $(a_2^\hcs-a_2^\st)$
is of the order of $20$ per cent, roughly an order of magnitude larger than the one for the highly dissipative cases of Fig.\ \ref{fig2}. Therefore, in order to obtain a good agreement between theory
and simulation (solid line), we have to insert into  \eqref{2.13} both the measured value of $B$ and the simulation value of the excess kurtosis difference $(a_2^\hcs-a_2^\st)$ \cite{difference}.
A similar situation, in which not only $B$ but also the excess kurtosis had to be taken from the simulations, was found in the analysis of the universal reference state of Ref. \cite{GMyT12} in
the same range of inelasticities.

\section{Sign and magnitude of the extremum}\label{sec:sign_and_mgnitude}

\subsection{Physical origin of the effect}
\label{ssec:physical_mechanism}

We attempt here a more physical explanation of the mechanism at work
here, which is, expectedly, very different from that in glassy
systems.  In essence, the effects we observe are subtle consequences
of energy dissipation, Without loss of generality, we focus on the
cooling protocol.  An important feature is the shape of the velocity
distribution $f({\vv},t)$, through the sign of the excess kurtosis
$a_2$. Is it ``flatter'' than the Gaussian (so-called platykurtic,
with $a_2<0$), or is it ``thinner'' (so-called leptokurtic, with
$a_2>0$) ?  {\em Distributions with $a_2<0$ dissipate less energy}\/
(and conversely, more energy when $a_2>0$).  Indeed, one can show that
to linear order in the excess kurtosis,
\begin{equation}
\frac{\langle v_{12}^n \rangle }{\langle v_{12}^n \rangle_0} \,=\, 1  \, + \,
a_2 \frac{n(n-2)}{16},
\label{eq:moment}
\end{equation}
where the average with index 0 refers to a Gaussian distribution of
the same variance, and $v_{12}$ is the modulus of the relative
velocity. The correction to unity vanishes when $n=0$ (normalization)
and $n=2$ (equality of variances).  Energy dissipation is related to the
moment $n=3$ (one $v$ coming from the collision frequency, and a $v^2$
from the fact that we are interested in the kinetic energy).  Thus
$\langle v_{12}^3 \rangle <\langle v_{12}^3 \rangle_0 $, for $a_2<0$
\cite{rque50}.

We start by discussing the behavior of the system in the
  cooling protocol, see Fig.~\ref{fig:protocol} (a), in which the
  driving in the waiting time window is smaller than the initial one,
  $\xi_{1}<\xi_{0}.$ Moreover, and for the sake of simplicity, we
  focus in the limiting case $\xi_{1}=0$, in which the system freely
  cools for $0<t<t_{w}$. We analyze the case $\xi_{1}\neq 0$ in
  Sec.~\ref{ssec:optimaltw}, in which we show that this change only
  affect the magnitude of the effect, but not its sign. Close to elasticity, $a_2<0$, for both driven and undriven gases
(platykurtic behavior).  It is quite difficult to shape an intuition
for the sign. It may be tempting to argue that it is a means for the
system to minimize energy dissipation, in spite of the lack of a general
principle holding for such non-equilibrium systems.  What is more
intuitive is that the unforced system shows stronger non Gaussianities
than the driven one, which benefits from stochastic kicks from the
forcing, {$|a_{2}^{\hcs}|/|a_{2}^{\st}|>1$ }.  Hence, at $t=t_w$, the system is in a state where $a_2$ is
more negative than it asymptotically will be, and therefore, energy
dissipation is, transiently, less.  This implies that $T$ shows a
maximum (or $\beta$ a minimum, as we observe).

The above scenario applies as long as dissipation is not too large
($\alpha >\alpha_{c}=1/\sqrt{2}$). On the other hand, for
$\alpha<\alpha_{c}=1/\sqrt{2}$, the driven and undriven systems become
leptokurtic ($a_2>0$, in order, in a hand-waving fashion, to cope with
large dissipation). We can subsequently follow the same reasoning as
above, which explains the anomalous effect.  The undriven kurtosis is
larger than the driven one (the driven $f$ is always the most
Gaussian), so that the larger value of $a_2$ at $t_w$ brings extra
dissipation. Thus, $T$ shows an undershooting (maximum of $\beta$).

For heating protocols, see Fig.~\ref{fig:protocol}(b), we next focus
on the limiting case $\xi_{1}\to\infty$. Again, a finite value of the
driving in the waiting time window $\xi_{1}$ does not change the sign
of the effect but only its magnitude, see next section. For a very
large value of $\xi_{1}$, the system rapidly evolves to a gaussian
distribution with $a_{2}=0$ in the waiting time window. Therefore, we
always have that $|a_{2}^{\st}|>|a_{2}^{\ini}|=0$ and following the
same line of reasoning as in the cooling case, it is easily shown that
the separation of the temperature from its steady value is simply
reversed.

{The above picture remains valid for a closely related
  thermostat, in which the energy injection is the same but the bath
  provides an additional friction force \cite{CVyG13}. In
  particular, the value of the excess kurtosis for that thermostat
  also verifies that $|a_{2}^{\st}|<|a_{2}^{\hcs}|$. The introduction
  of this additional friction force allows the system to reach a
  well-defined steady state even in the elastic limit $\alpha=1$, in
  which the dissipation stemming from collisions disappears.}

\subsection{The optimal waiting time}
\label{ssec:optimaltw}

\begin{figure}
  \centering
  \includegraphics[width=3.25in]{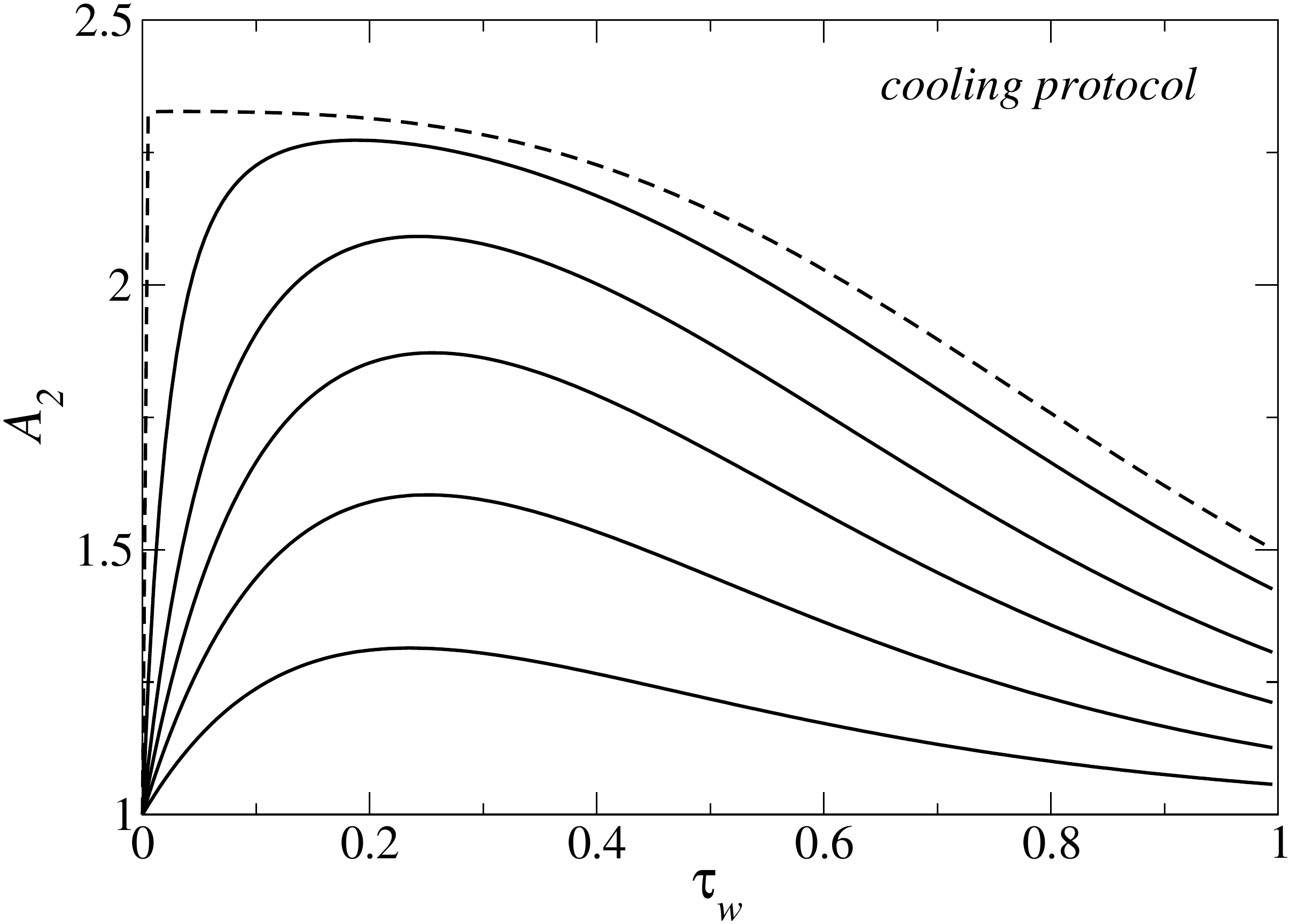}\\
  \caption{Evolution of excess kurtosis ratio, $A_2(\uptau_w)\equiv
    a_2(\uptau_w)/a_2^\st$, as a function of waiting time, within the
    cooling protocol at $\alpha=0.3$. From bottom to top, the curves
    correspond to $T_s(\xi_0)/T_s(\xi_1) = 2, 4, 9, 25$ and 200. The
    upper dashed curve is for the limit $T_s(\xi_1)/T_s(\xi_0)\to
    0$. Note that $A_2(\uptau_w)$ defines the quantity $A_2^\ini$ used
    throughout. For a given value of $\alpha$, the maximum possible
    $A_2$ is $a_2^\hcs/a_2^s$. For $\alpha=0.3$,
    Fig. \ref{fig:a2ratio} indicates that this ratio is close to $2.33$,
    which is consistent with the maximum of the dashed curve.}
  \label{fig:waiting_cool_alpha0.3}
\end{figure}

\begin{figure}
  \centering
  \includegraphics[width=3.25in]{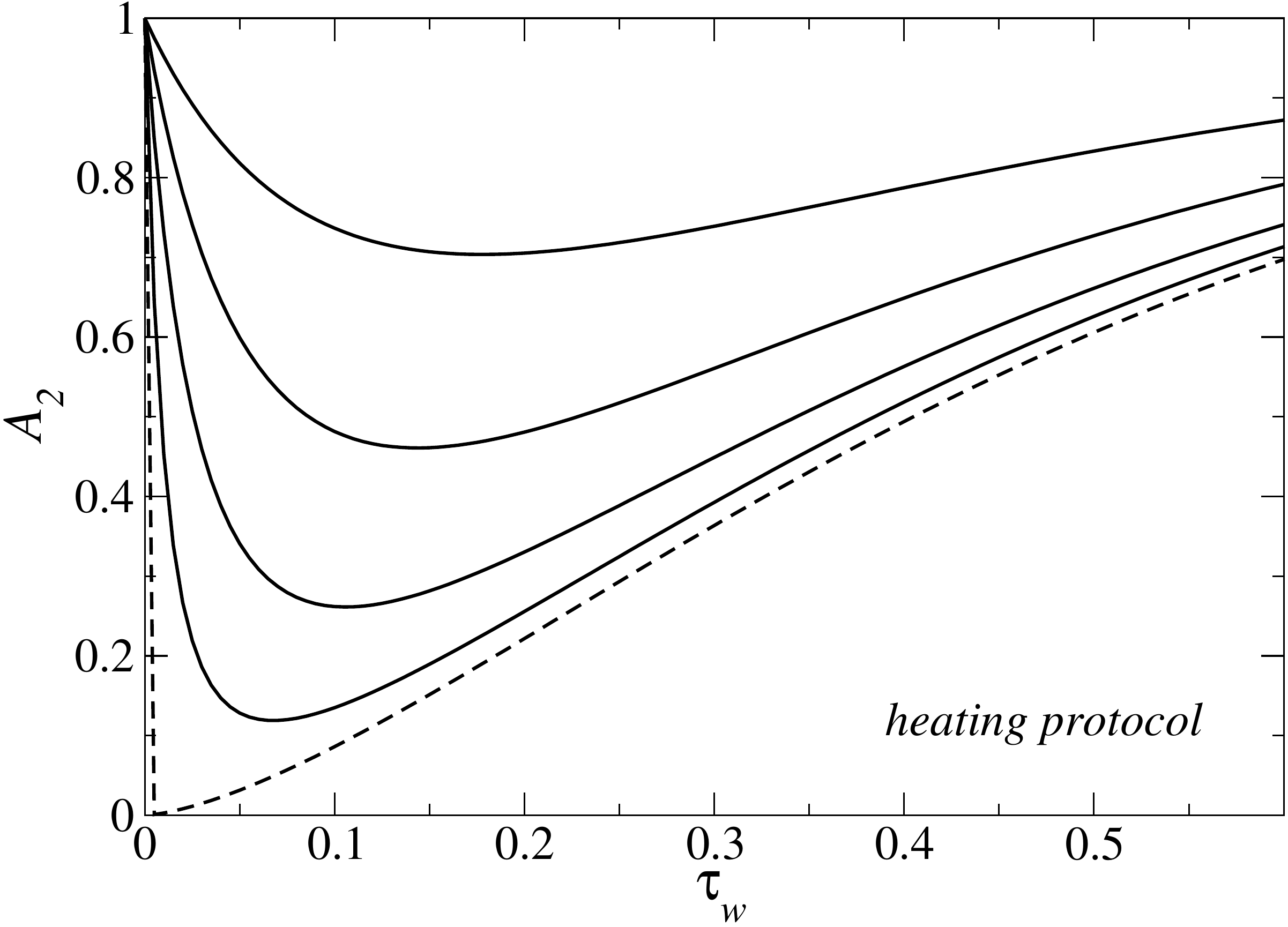}\\
  \caption{Same as Fig. \ref{fig:waiting_cool_alpha0.3} but for the heated protocol.
  Here, from top to bottom: $T_s(\xi_1)/T_s(\xi_0)=2, 4, 9, 25$. The lower dashed curve
  is for $T_s(\xi_1)/T_s(\xi_0)\to \infty$}
  \label{fig:waiting_heat_alpha0.3}
\end{figure}

\begin{figure}
  \centering
  \includegraphics[width=3.25in]{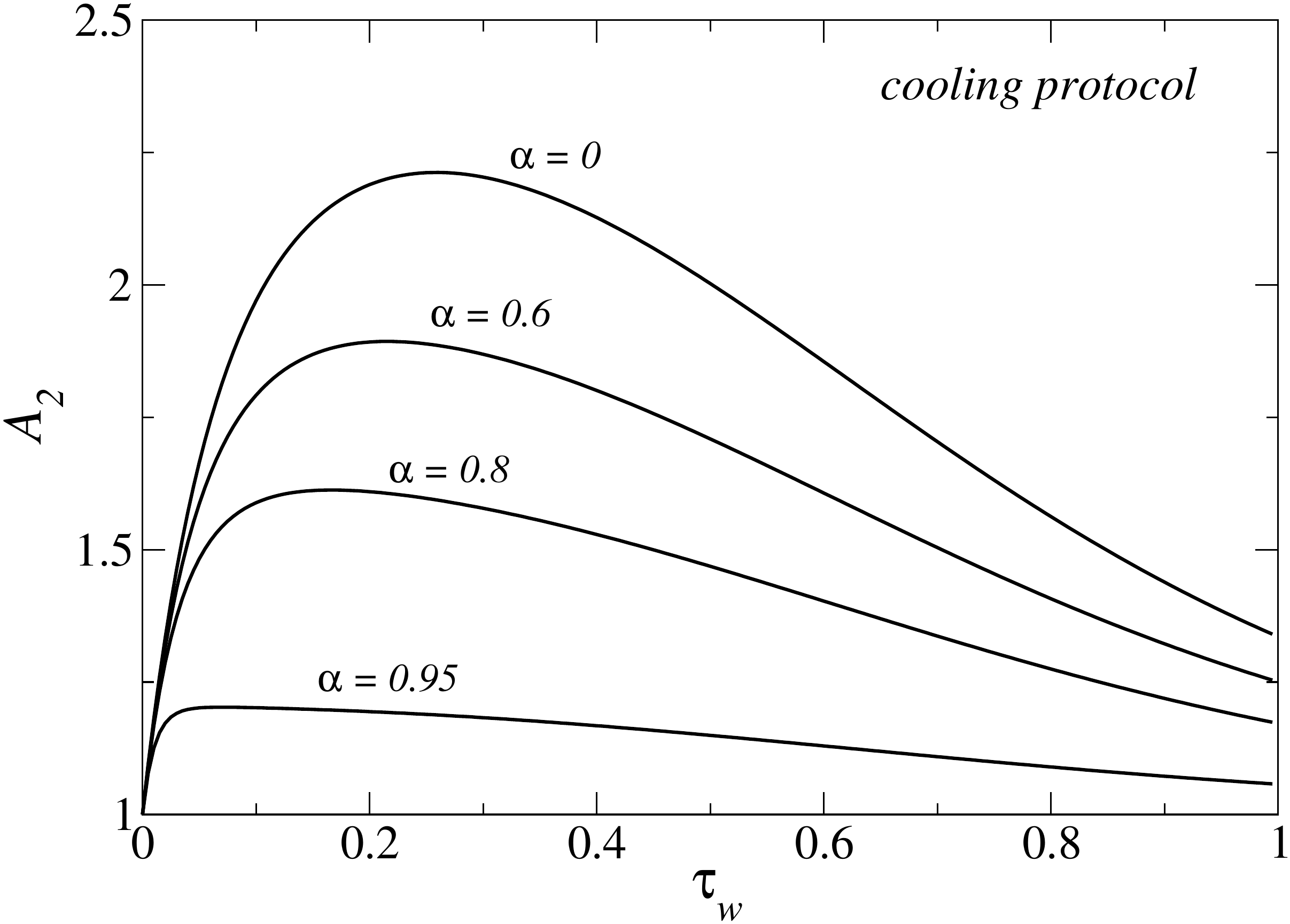}\\
  \caption{Excess kurtosis ratio as a function of waiting time (cooling protocol),
  for different dissipations, and $T_s(\xi_1)/T_s(\xi_0)=1/25$.}
  \label{fig:waiting_cool_alpha_all}
\end{figure}

We now return to the cooling protocol, in the limiting case where
$\xi_1/\xi_0$ is close to zero. At $\xi_1/\xi_0=0$, the waiting time
$t_w$ can be arbitrarily large, since $a_2$ will evolve to $a_2^\hcs$,
and the longer one waits (in real time scale, not in the $\tau$ scale,
see below), the stronger the effect.  In general however, there is an
optimal value of $t_w$, which depends on the ratio
$T_s(\xi_1)/T_s(\xi_0)$, for which the amplitude of the Kovacs
response is maximal. The reason is that the difference in kurtosis,
$|a_2(t_w)-a_2^{\text{s}}|$, should be maximized.  If one spends too
much time in the waiting window, the system can attain its
non-equilibrium steady state, $a_2(t_w)$ then reaches the value
$a_2^{\text{s}}$ ($A_2\to 1$), and the humps disappears.  This holds
for both the cooling ($\xi_1<\xi_0$) and the heated ($\xi_1>\xi_0$)
protocols, see Figures \ref{fig:waiting_cool_alpha0.3},
\ref{fig:waiting_heat_alpha0.3} and \ref{fig:waiting_cool_alpha_all}.
These figures therefore exhibit an extremum at a particular value of
$\uptau_w$, which provides the optimal waiting time. It can be
observed that in the $\uptau$ scale, this optimum depends only weakly
on $\xi_1/\xi_0$ (or equivalently on $T_s(\xi_1)/T_s(\xi_0)$), and
likewise, quite weakly on dissipation.

The trends observed in the
Figures, with a maximum (resp.~minimum) in the cooling (resp.~heating)
case, can be understood as in Sec.~\ref{ssec:physical_mechanism}, and are fully consistent with the
argument put forward there. 
In the extreme case $T_s(\xi_1)/T_s(\xi_0) \to \infty$ (that is,
$\xi_1/\xi_0\to\infty$), the velocity distribution is provided enough
time to become Gaussian, with thus a vanishing $a_2$ (and $A_2$). This
is the behavior shown in Fig.  \ref{fig:waiting_heat_alpha0.3}. Yet,
the dashed line also shows that for any finite
$T_s(\xi_1)/T_s(\xi_0)$, no matter how large, the optimal waiting time
becomes vanishingly small in the $\tau$ scale, which reflects the fact
that under extreme forcing $\xi_1$, the system is so much driven that
it is able to quickly reach its steady-state. It is at this point
interesting to turn to the dashed line in
Fig. \ref{fig:waiting_cool_alpha0.3} for the cooled extreme case
$\xi_1/\xi_0\to 0$. It also reveals that the optimal $\tau_w$ also
vanishes, whereas, on intuitive grounds, it should be that one can
wait arbitrarily long without seeing the system depart from the
homogeneous cooling state it quickly attains. In other words, one may
expect that the optimal waiting time should diverge upon decreasing
the forcing.  This is the case, but it can only be appreciated by
returning to the original $t$ scale: it turns out that the optimal
$t_w \propto \tau_w/\sqrt{T_s(\xi_1)}$ diverges when $\xi_1 \to 0$,
due to the vanishing of $T_s(\xi_1)$.

\begin{figure}
  \centering
\includegraphics[width=3.25in]{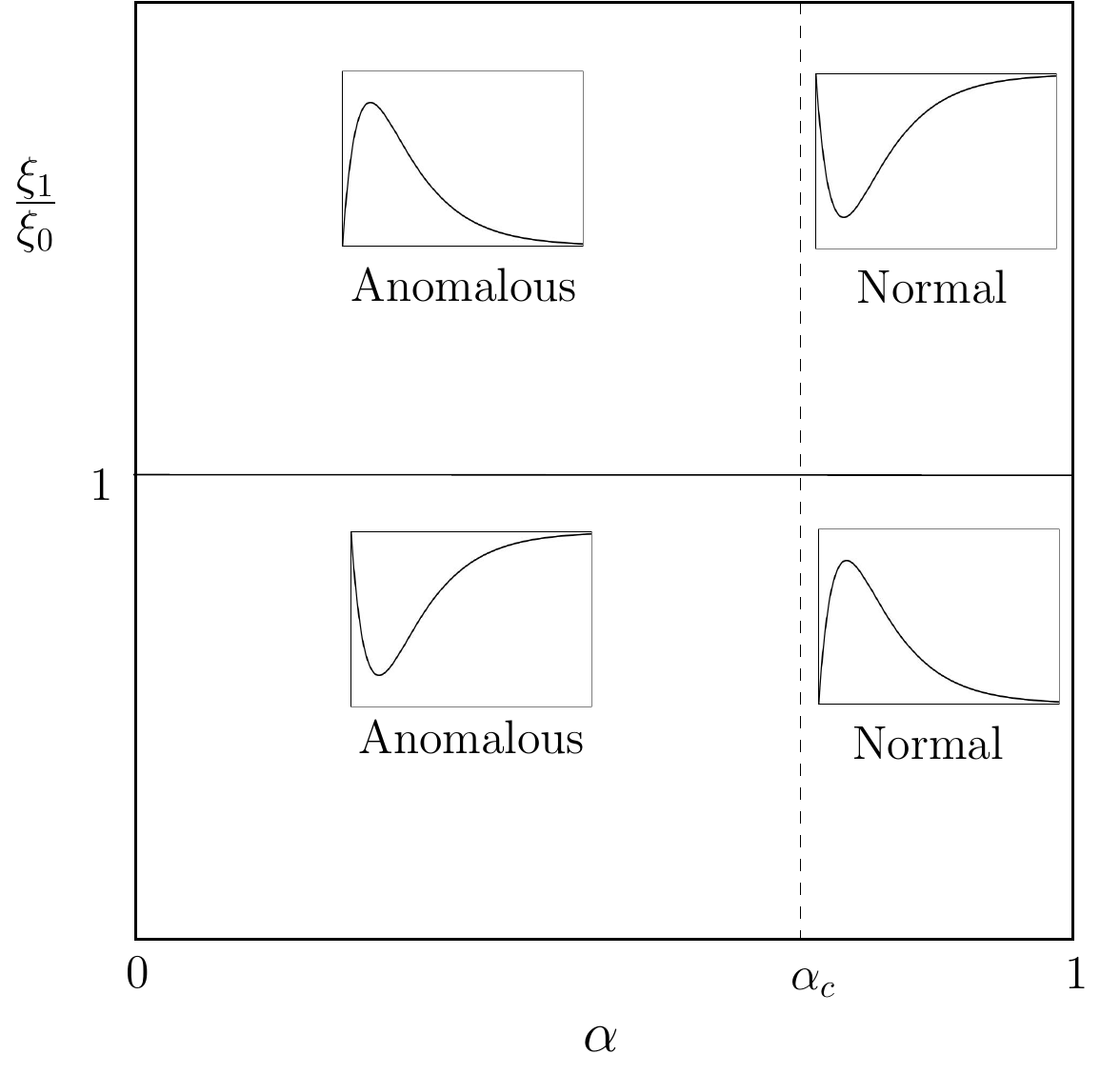}
  \caption{\label{phase_diag} Phase diagram of the Kovacs hump. The
    line $\xi_{1}/\xi_{0}=1$ (solid) separates the ``cooling''
    ($\xi_{1}<\xi_{0}$) and the ``heating'' ($\xi_{1}>\xi_{0}$)
    protocols. The dashed line $\alpha=\alpha_{c}=1/\sqrt{2}$
    separates systems with ``high inelasticity''
    ($\alpha<\alpha_{c}$) from those with ``low inelasticity''
    ($\alpha>\alpha_{c}$). {Note that the plots are for the
      granular temperature $T$, a maximum in $T$ corresponds to a minimum
      in the $\beta$ variable defined in Eq.~\eqref{1.12}}. }

\end{figure}

{We attempt here a summary of the main results reported in this Section.
The Kovacs-like protocol used throughout this paper can be described
  by three dimensionless parameters: (i) the restitution coefficient
  $\alpha$, (ii) the ratio $\xi_{1}/\xi_{0}$ of the intermediate driving $\xi_{1}$ to
  the initial one $\xi_{0}$,
  and (iii) the dimensionless waiting time $\uptau_{w}$, which in turn
  fixes the ratio $\xi/\xi_{1}$. The sign of the hump is completely
  determined by the first two, $\alpha$ and $\xi_{1}/\xi_{0}$, while
  the third only affects the magnitude of the extremum. A
  \textit{phase diagram} of the Kovacs hump is sketched in
  Fig.~\ref{phase_diag}. The ``normal'' behavior is similar to the one
  observed in molecular systems when controlling the bath temperature
  and measuring the energy (or the volume). The lines in the diagram
  indicate the values of the parameters for which no Kovacs hump would
  be observed. The solid line $\xi_{1}=\xi_{0}$ separating heating and
  cooling protocols delineates a ``trivial'' boundary, with no change in the
  driving and thus no hump. On the other hand, the dashed line
  $\alpha=\alpha_{c}$ separating the low and high inelasticity regions
  is less expected, and follows from the accurate prediction of the first Sonine
  approximation for the change of sign in the Kovacs hump.}

\section{Long time behavior and compatibility with the universal reference state}
\label{sec3b}

On close inspection, the trends reported above for the time evolution
of $\beta$ are not compatible with the requirement that the system
should asymptotically evolve towards the universal state brought to
the fore in Ref. \cite{GMyT12}. We discuss and resolve that question
here. In a nutshell, the time evolution is slightly more complex than
the simplified expressions obtained in Section
\ref{ssec:analytical}. For the sake of simplicity, we use in
  this section the shifted time variable
  $\uptau=\zeta_0\sqrt{T_s}(t-t_w)/2$, which vanishes at $t=t_{w}$.
Let us consider the equation for the first-order correction to the
excess kurtosis,
\begin{equation}\label{2.14}
  \frac{dA_{21}}{d\uptau}+4BA_{21}=-\left[ (12-4B) A_{20}+ 4B \right] \beta_1.
\end{equation}
We do not write here its complete solution, but only its leading behavior for long times. The solution of \eqref{2.14} is  a linear combination of exponentials with different relaxation times.
For $\uptau\to\infty$, the rhs of  \eqref{2.14} behaves, to dominant order, as
\begin{equation}\label{2.14bis}
  h(\uptau)=-12\gamma \Delta A_2^\ini e^{-3\uptau} ,
\end{equation}
as follows from Eq. \eqref{2.8b} and (\ref{2.11}).
The term in $A_{21}$ coming therefrom is
\begin{equation}\label{2.15}
  A_{21}^h(\uptau)=-64 \gamma^2 \Delta A_2^\ini e^{-3\uptau},
\end{equation}
and asymptotically dominates
\begin{equation}\label{2.16}
  A_{21}(\uptau)\sim A_{21}^h(\uptau), \quad \uptau\gg 1.
\end{equation}
Interestingly, this term is much bigger than $A_{20}(\uptau)$ for very long times, and thus gives the long time tendency to the steady value of the rescaled excess kurtosis,
\begin{equation}\label{2.17}
  A_2(\uptau)-1 \sim a_2^\st A_{21}^h(\uptau), \quad \uptau\gg 1.
\end{equation}
The condition for the asymptotic result in \eqref{2.17} to hold is,
more concretely, $\exp(-4B\uptau) \ll \exp(-3\uptau)$ or,
equivalently, $\exp[-(4B-3)\uptau]\ll 1$. It is worth noting that the
sign of $A_{21}^h(\uptau)$ is opposite to that of $\Delta A_2^\ini$
and therefore different from that of the zero-th order contribution
$A_{20}(\uptau)-1$, see Eq. (\ref{2.8b}).  As $a_2^\st<0$ for weakly
dissipative systems, $\alpha>\alpha_c$ while $a_2^\st>0$ in the highly
dissipative case, $\alpha<\alpha_c$, Eq. \eqref{2.17} predicts that,
for long times $\uptau\gg 1$, the sign of $A_2-1$ is the opposite to
that of $A_{20}-1$ for $\alpha<\alpha_c$. This means that $A_2$ has a
minimum and tends to unity from below in the highly dissipative case.
This behavior was overlooked by the analysis performed in previous
sections. The effect is quite small and thus difficult to measure in
the simulations, but it has important theoretical consequences.  In
Ref.~\cite{GMyT12} it was proved that, for long enough times, a
uniformly heated granular gas reaches the \textit{universal reference
  $\beta$-state}, over which all the time dependence can be encoded in
$\beta$. In other words, for long enough times, all the moments of the
velocity distribution function (for instance, the excess kurtosis)
forget their initial conditions and become only a function of the
``distance'' $\beta$ to the steady state. Afterwards, for even longer
times, $\beta$ approaches its steady value. For the excess kurtosis,
and in the linear regime close to the steady state, this universal
behavior is given by
\begin{equation}\label{2.18}
  A_2-1 \sim\left. \frac{dA_2}{d\beta}\right|_{\beta=1}(\beta-1)=-\frac{12}{4B-3}(\beta-1).
\end{equation}
The value of the derivative $dA_2/d\beta|_\beta=1$ has been calculated by applying L'H\^opital rule to Eq. (19) of Ref.~\cite{GMyT12}.

If we take the lowest order approximation for both $A_2-1$, which is $A_{20}-1$, and for $\beta-1$, which is given by $\beta_1$, we have that
\begin{equation}\label{2.19}
  \lim_{\uptau\to\infty} \frac{A_{20}-1}{\beta-1}=0,
\end{equation}
in strong disagreement with \eqref{2.18}, which predicts a value
$-12/(4B-3)<0$ instead.  This problem is mended if we consider, as
should be done, $A_2-1$ and $\beta-1$ up to the same order. Since the
dominant term for long times in the decay of $A_2$ is proportional to
$A_{21}^h$, as given by \eqref{2.17}, and the long time behavior of
$\beta-1$ can be straightforwardly inferred from \eqref{2.13},
\begin{equation}\label{2.20}
  \beta(\uptau)-1 \sim a_2^\st \gamma \Delta A_2^\ini  e^{-3\uptau},
\end{equation}
one obtains that
\begin{equation}\label{2.21}
  \frac{A_2-1}{\beta-1}\sim -64\gamma \,=\, \frac{-12}{4B-3}  , \quad \uptau\gg 1,
\end{equation}
where the definition of $\gamma$, Eq. \eqref{2.12}, has been used.
The result in \eqref{2.21} is in agreement with \eqref{2.18}.

\begin{figure}
  \centering
  \includegraphics[width=3.25in]{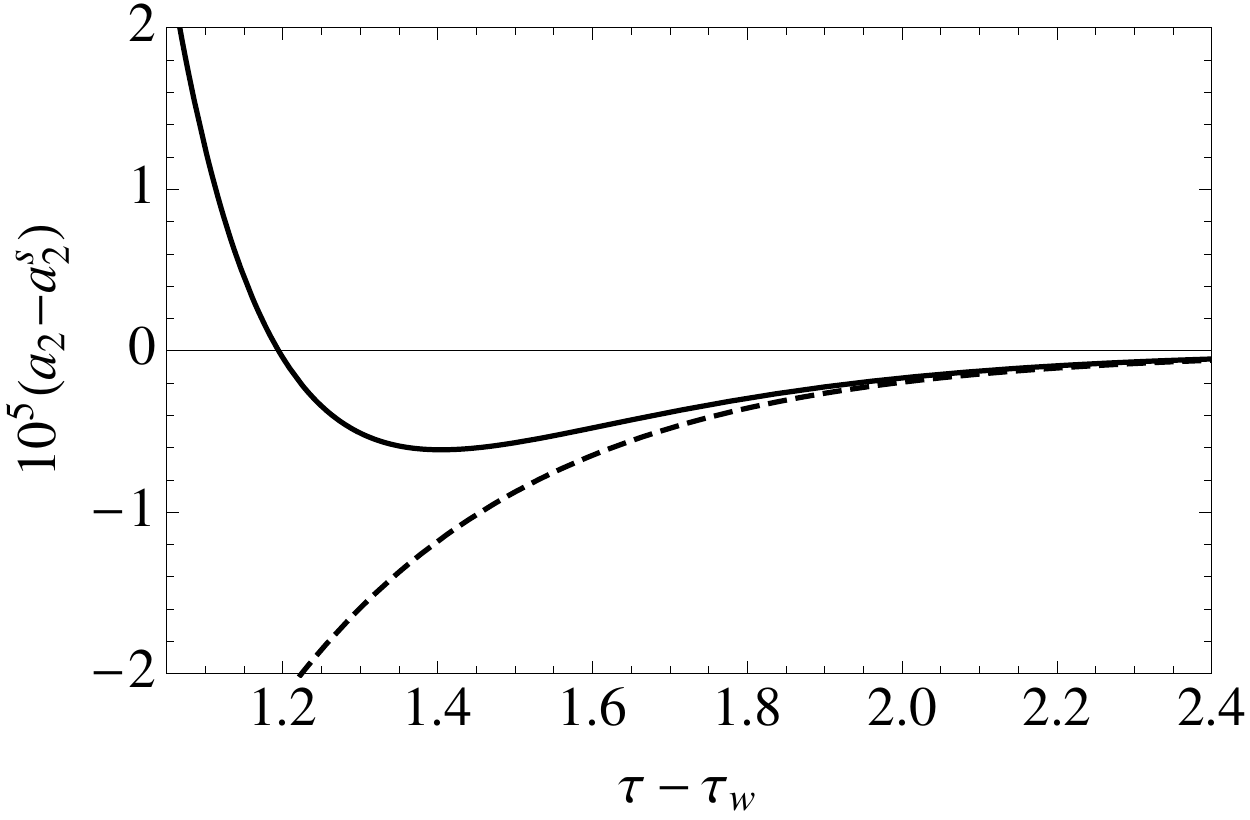}\\
  \caption{Tendency to the universal reference state for very long
    times. We show a zoom of the long time behavior
    ($\uptau-\uptau_{w}\geq 1$) of the decay of the excess
    kurtosis to its steady value, $|a_2-a_2^\st|\leq 2 \times
    10^{-5}$. The overall picture is that of Fig. \ref{fig:a2decay},
    which also corresponds to $\alpha=0.3$, for which
    $a_2^\ini-a_2^\st \simeq 0.086$. Plotted here is the excess
    kurtosis decay obtained from (i) the numerical integration  of
    Eq. \eqref{1.17} with initial conditions \eqref{2.2} (solid line)
    (ii) the asymptotic behavior given by Eq. \eqref{2.17} and
    \eqref{2.15} (dashed line).
    }\label{fig:reference}
\end{figure}

Figure \ref{fig:reference} shows the tendency of the system to
approach the universal reference state for very long times. Although
to the zero-th order the overall relaxation of the excess kurtosis to
the steady state is very well described by a single exponential, see
Fig. \ref{fig:a2decay}, for very long times $a_2-a_2^\st$ changes sign
and tends to zero from below.  This is in full agreement with the
approach to the universal reference state, as described by
Eq. \eqref{2.18} or \eqref{2.21}. The minimum is tiny, being four
orders of magnitude smaller than the initial distance to the steady
state for the plotted case ($\alpha=0.3$). This makes it very
difficult to measure this effect in simulations. However, it is
crucial from a theoretical point of view, since it shows that the
theoretical approach developed here is compatible with the general
long time behavior derived in Ref. \cite{GMyT12}.

\section{Final remarks}
\label{sec4}

In conclusion, we have studied from a granular gas perspective a
memory effect that pertains to glassy phenomenology. A striking
consequence of the analysis is that the sign of the Kovacs hump
changes as the restitution coefficient is varied from the
quasi-elastic limit $\alpha\to 1^-$ to the completely inelastic case
$\alpha=0$. There is a critical value of the restitution coefficient
$\alpha_c$, which coincides with the point at which the stationary
value of the excess kurtosis changes sign. First, we recapitulate the
behavior for cooling protocols as the one depicted in
Fig.~\ref{fig:protocol}(a). For weakly dissipative systems, in the
sense that $\alpha>\alpha_c$, the granular temperature passes through
a maximum, larger than its corresponding steady value $T_{\st}$
($\beta=\sqrt{T_\st /T}<1$). The sign of the hump changes for highly
dissipative systems, in which $\alpha<\alpha_c$: the temperature
passes through a minimum ($\beta>1$). Conversely, for heating
protocols, in which $\xi_0<\xi<\xi_1$ as sketched in
Fig.~\ref{fig:protocol}(b), we simply have a reversal of the sign of
the hump: the granular temperature displays a minimum for small
inelasticity, $\alpha>\alpha_{c}$ and a maximum for high inelasticity
$\alpha<\alpha_{c}$. Table \ref{table:summary} summarizes the
phenomenology.  On the other hand, in a molecular system, the measured
quantity in the analogous experimental situation \cite{molecular}
always exhibits a maximum (resp.~minimum) in the cooling
(resp.~heating) protocol. This stems from the mathematical structure
of the analytical expression for the Kovacs hump within linear
response theory, but the same result seems to remain valid in the
nonlinear regime \cite{PyB10,DyH11,RyP14}.

\begin{table*}
\begin{center}
  \begin{tabular}{|c||c|c|c|c|c|c|c|}
\hline
   protocol & inelasticity & $\alpha$ & $a_{2}^{\ini}-a_{2}^{\st}$ & dissipation &   $T$ hump & Kovacs effect \\
\hline
\hline
   cooling &  ``low''& $\; >\alpha_{c} \;$ & $\; <0 \;$   & smaller than stationary  &   maximum  & normal\\
 cooling & ``high''& $\; <\alpha_{c} \;$ & $\; >0 \;$   & larger than stationary  &   minimum  & anomalous\\
\hline
   heating &  ``low''& $\; >\alpha_{c} \;$ & $\; >0 \;$ &  larger than stationary  &  minimum   & normal\\
 heating & ``high''& $\; <\alpha_{c} \;$ & $\; <0 \;$ &  smaller than stationary  &   maximum  & anomalous\\
\hline
\end{tabular}
\caption{\label{table:summary} Hump phenomenology and
    the underlying physical mechanism for the cooling and heating driving protocols in
    Fig.~\ref{fig:protocol}. The `critical' value of the restitution
    coefficient $\alpha$ is $\alpha_c=1/\sqrt{2}$.}
\end{center}
\end{table*}

Therefore, the Kovacs effect for uniformly heated granular gases is
\textit{normal} for small inelasticities while it is
\textit{anomalous} in the highly inelastic case, independently of the
details of the protocol followed in the waiting time window. The
intermediate value of the driving $\xi_{1}$ and the waiting time
$t_{w}$ do affect the amplitude of the memory effect, but not its sign
and shape, as expressed by Eq.~\eqref{factorized} and discussed in
Sec.~\ref{sec:sign_and_mgnitude}. Nevertheless, there are optimal
values of $\xi_{1}$ and $t_{w}$ that maximize the amplitude of the
hump for a given value of the restitution coefficient.  Quite
intuitively, for the usual cooling protocol the optimal choice of
parameters corresponds to the limit $\xi_{1}\to 0$ with a large enough
$t_{w}$, such that the system ends up in the homogeneous cooling state
inside the waiting time window.

In molecular systems, energy is conserved and, within the linear
response regime, the shape of the Kovacs hump is closely related to
the linear relaxation function of the energy from the initial
temperature $T_{0}$ to the final one $T$. This \textit{direct}
relaxation function decays monotonically because it is proportional to
the equilibrium time autocorrelation function of the energy, as stated
by the fluctuation-dissipation theorem \cite{vK92}. In turn, this
monotonicity assures that the Kovacs hump is always positive for the
usual cooling protocol \cite{PyB10}, while it is negative for the
heating protocol considered in Ref.~\cite{DyH11}. Therefore, it seems
worth investigating the anomalous character of the Kovacs hump found
here for high dissipation. {Specifically}, it would be
interesting to analyze the possible relation between the anomalous
character of the Kovacs effect for high dissipation and the validity
of the fluctuation-dissipation relation in {non-equilibrium
  systems. In the context of granular media, there is some recent work
  trying to establish the validity of fluctuation-dissipation
  relations.}
\cite{PByL02,PByV07,MGyT09,SVGP10,PLyH11-12,BMyG12}. {It seems
  particularly appealing to investigate simple models of dissipative
  systems \cite{PLyH11-12,BByP02}, for which the calculations may be
  carried out without introducing any approximations like the Sonine
  expansion considered here.}

Our main assumptions are (i) the accurateness of the first-Sonine
approximation (ii) the smallness of the excess kurtosis that makes it
possible to neglect nonlinear terms in $a_2$. Our expression for the
Kovacs hump, as given by Eq. \eqref{2.13}, is valid up to the linear
order in the excess kurtosis. If nonlinear corrections in $a_2$ were
incorporated to the time evolution equations, this linear order result
would not be affected. The exponential decay of the excess kurtosis to
the zero-th order, as given by $A_{20}$, is neither affected by the
introduction of nonlinearities. The same is applicable to the long
time behavior and the tendency to the universal reference state
discussed in Sec. \ref{sec3b}. This may be surprising at first sight,
because nonlinearities in $a_2$ should certainly change the equation
for the excess kurtosis first-order correction $A_{21}$. However,
these nonlinearities must vanish in the steady state (as $(A_2-1)^2$
to the quadratic order), and thus they are subdominant against the
leading term as given by $h(\uptau)$, Eq. \eqref{2.14bis}. The results
derived throughout the paper are therefore robust.

One of the main implications of the original work by Kovacs is that it
clearly showed that the experimental macroscopic variables (pressure,
volume, temperature, for polymers) do not suffice to completely
characterize the system state, which in general depends on the whole
previous thermal history. In this sense, the existence of the Kovacs
hump here, independently of its amplitude and sign (normal or
anomalous), is a crisp proof that the state of the uniformly heated
granular gas is not uniquely determined by its granular temperature,
and other variables must be incorporated to have a complete
description thereof. At first glance, this conclusion seems similar to
that reached in the analysis of its universal reference $\beta$-state
\cite{GMyT12,GMyT13}, in which it was shown that the ``distance'' to
the steady state $\beta$ is also necessary to describe the uniformly
driven granular gas.  But it must be stressed that here, we go
further. While the $\beta$-state reached for long times is uniquely
determined by the driving $\xi$ and the granular temperature $T$, we
show the relevance of explicitly keeping track of the intrinsic
dynamics of non-Gaussianities, through the decoupling of $a_2$ and
$\beta$.

In principle, a similar behavior should appear for other kinds of
drivings, provided that the driving intensity and the granular
temperature do not suffice to completely characterize the state of the
system. Within the first Sonine approximation, the magnitude of the
Kovacs hump would be proportional to the difference between the
initial value of the excess kurtosis $a_{2}^{\ini}$ and its steady
value for the considered thermostat \cite{other_driving}.  In the usual cooling protocol,
if a very low value of the intermediate driving $\xi_{1}$ were used,
the value of the excess kurtosis after the waiting time
would be close to that of the homogeneous cooling
  state. Therefore, non-Gaussianities are a necessary but not
sufficient condition to have  memory effect of the kind reported here in a driven
granular gas \cite{gaussian_thermostat}. In all generality, the
possibility of having a transition from normal to anomalous Kovacs
effect is encoded in the change of sign of $a_2^\ini-a_2^\st$.

The Kovacs hump in granular gases occurs over the kinetic time
scale. For the time at which the temperature passes through its
extremum, the system has not reached the \textit{hydrodynamic} stage
\cite{Re77} in which the all the time dependence of the
  velocity distribution function occurs through the hydrodynamic
  fields (density, average velocity and temperature), and initial
conditions have been forgotten. Over the hydrodynamic $\beta$-state of
uniformly driven gases, the decay of the temperature (or of $\beta$)
to its steady value is a monotonic function of time
\cite{GMyT12,GMyT13}. Here, this monotonicity condition is only
fulfilled for times greater than that of the extremum. Then, the
system reaches this hydrodynamic solution of the Boltzmann equation
only for very long times, when it is linearly close to the steady
state.

\acknowledgments

We acknowledge useful discussions with M.I. Garc\'{\i}a de
Soria and P. Maynar. This work has been supported by the Spanish
Ministerio de Econom\'\i a y Competitividad grant FIS2011-24460 (AP).
AP would also like to thank the Spanish Ministerio de Educaci\'on,
Cultura y Deporte mobility grant PRX12/00362 that funded his stay at
the Universit\'e Paris-Sud in summer 2013, during which this work was
carried out.


\begin{thebibliography}{99}
\bibitem{P98a}
A slight variant of the model can be found in
\cite{P98b,CVyG13,PSD13}.
\bibitem{typo}
There is a typo in the expression for $B$ of Ref.~\cite{GMyT12}, concretely in the sign of the term in the denominator proportional to $(62-30\alpha)$, which has been corrected upon writing Eq.~\eqref{1.15}.
\bibitem{rque88}
It can be noted that under the stochastic forcing with drag studied
in Ref. \cite{CVyG13}, the excess kurtosis does also change sign
at $\alpha=1/\sqrt{2}$, keeping a functional dependence on $\alpha$
that is close to that considered here.
\bibitem{long_enough_tw}
A long enough $t_{w}$ is easily attained by starting from a high
enough initial value of the driving $\xi_{0}$, that is, a high enough
granular temperature.
\bibitem{difference}
For $\alpha=0.8$, the theoretical estimates of the excess kurtosis are $a_2^\hcs=-0.02243$ and $a_2^\st=-0.01349$, so that $a_2^\hcs-a_2^\st=-0.00895$, while the simulation values are $a_2^\hcs=-0.02635$ and $a_2^\st=-0.01495$, which lead to $a_2^\hcs-a_2^\st=-0.01140$.
\bibitem{rque50} Note however that for the moment $n=1$ (related to
  the collision frequency), the inequality is reversed.  The change of
  sign of the correction, between $n=1$ and $n=3$, illustrates the
  subtleness of the effect. Platykurtic shapes exhibit depleted
  distributions for small velocities, then enhanced population around
  the thermal scale, and again depletion for slightly larger
  velocities (not speaking about the truly large velocity tail, which
  does not matter here, and which is overpopulated
  \cite{NyE98,BT03}). It is the balance of these over/under
  populations that leads to Eq. (\ref{eq:moment}) above.  Note also
  that Eq. (\ref{eq:moment}) explains the presence of the contribution
  $3 \,a_2/16$ in Eq. (\ref{1.7}), see also \cite{NyE98}.
\bibitem{molecular} Let us remember that, in molecular systems, the
  role of the granular temperature is usually played by the volume or
  the energy, while the role of the driving is played by the bath
  temperature. The steady value of the
    granular temperature is an increasing function of the driving, a
    trend that is similar to the increase of the equilibrium value of
    the energy or volume with increasing bath temperature.
\bibitem{other_driving}
This has been very recently checked, see for instance
Eq.~(37) of Ref.~\cite{BGMyB14}, in which this kind of memory effect
has been analyzed for a different thermostat, within the first Sonine approximation.
\bibitem{gaussian_thermostat}
The case of the so-called Gaussian thermostat would then be  special, because it can be mapped onto the homogeneous cooling state.
Thus, all the time dependence of the system is encoded in the temperature and,
in particular, the value of the excess kurtosis is known \cite{MyS00}.
\end{thebibliography}
\end{document}